\documentstyle[12pt,epsfig]{article}
\def \medio  {\baselineskip= 1.5 \normalbaselineskip}

\newcommand{\titul}[1] {\begin{center}{\large\bf #1 } \end{center}\vskip 1.cm}
\newcommand{\autor}[1] {\begin {center} {\large \lineskip .5em #1 }
                        \end   {center} }
\newcommand{\lugar}[1] {\begin{center} {\it #1} \end{center}}
\newcommand{\abstr}[1] {{\begin{center} \vskip .5cm {\bf Abstract
                        \vspace{0pt}} \end{center}}\begin{quote} #1
                        \end{quote}}
\newcommand{\vph}{\varphi}
\newcommand{\bea}{\begin{eqnarray}}
\newcommand{\eea}{\end{eqnarray}}
\begin{document}

\begin{titlepage}

\begin{flushright} {
%10th version \\
\bf US-FT/7-01 \\ 
June, 2001
} \end{flushright}

%\vskip 3.cm
\titul{ The contribution of off-shell gluons\\
to the structure functions $F_2^c$ and $F_L^c$\\
and the unintegrated gluon distributions
}

\autor{
A.V. Kotikov
%\footnote{E-mail:kotikov@sunse.jinr.ru}
}
\lugar{Bogoliubov Laboratory of Theoretical Physics\\
Joint Institute for Nuclear Research\\
141980 Dubna, Russia}
\autor{A.V. Lipatov}
\lugar{Department of Physics \\
M.V. Lomonosov Moscow State University\\
119899 Moscow, Russia}
\autor{G. Parente}
%\footnote{E-mail:gonzalo@fpaxp1.usc.es}
\lugar{Departamento de F\'\i sica de Part\'\i culas\\
Universidade de Santiago de Compostela\\
15706 Santiago de Compostela, Spain}
\autor{N.P. Zotov}
\lugar{D.V. Skobeltsyn Institute of Nuclear Physics \\
M.V. Lomonosov Moscow State University\\
119899 Moscow, Russia}
\abstr{
\medio
We calculate the perturbative parts of the structure functions
$F_2^c$ and $F_L^c$ for a gluon target having nonzero transverse
momentum squared at order $\alpha _s$.
The results of the double convolution (with respect to the Bjorken variable
$x_B$ and the
transverse momentum ) of the perturbative part and the unintegrated
gluon densities are compared with HERA experimental data for
$F_2^c$.
The contribution from $F_L^c$ structure function ranges $10\div30\%$
of that of $F_2^c$ at the kinematical range of HERA experiments.  

\vskip 0.5cm

PACS number(s): 13.60.Hb, 12.38.Bx, 13.15.Dk

}
\end{titlepage}
\newpage

\pagestyle{plain}
\medio
\section{Introduction} \indent 

Recently there have been important new data on the charm structure function
(SF) $F_2^c$, of the proton from the H1 \cite{H1,H1c} and ZEUS \cite{ZEUS,
ZEUS2} 
Collaborations at HERA, which have probed the small-$x_B$ region down to 
$x_B=8\times 10^{-4}$ and $x_B=2\times 10^{-4}$, respectively. At these values 
of $x_B$, the charm contribution to the total proton SF, $F_2^p$, is found
to be around $25\%$, which is a considerably larger fraction than that found 
by the European Muon Collaboration at CERN \cite{EMC} at 
%samewhat 
larger $x_B$, where it was only $\sim 1\%$ of $F_2^p$. Extensive 
theoretical analyses 
in recent years have generally served to confirm that 
%the bulk of 
the
$F_2^c$ data can be described through perturbative generation of charm 
within QCD (see, for example, the review in Ref. \cite{CoDeRo} and references 
therein).

In the framework of DGLAP dynamics \cite{DGLAP1,DGLAP} there are two basic
methods to study heavy flavour physics. One of them \cite{Kniehl} is based on 
the massless evolution of parton distributions and the other  
\cite{Flixione} on the boson-gluon
fusion process. There are also interpolating schemes (see Ref. \cite{Aivazis}
and references therein). The present HERA data \cite{H1,ZEUS, ZEUS2,H1c} 
for the charm SF $F_2^c$ are
in good 
agreement with the predictions from Ref. \cite{Flixione}.

We note, however, that perhaps more relevant analyses of the HERA data, where 
the $x_B$ values are quite small, are those based on BFKL dynamics
\cite{BFKL} (see discussions in the review of Ref. \cite{Kwien} and references
therein), because the leading $ln(1/x_B)$ contributions are summed. The basic
dynamical quantity in BFKL approach is the unintegrated gluon distribution
$\vph_g(x,k^2_{\bot})$ 
%$\Phi(x_B,k^2_t)$ 
($f_g$ is the (integrated) gluon distribution multiplied
by $x_B$ and $k_{\bot}$ is the transverse momentum)
 \begin{eqnarray}
f_{g}(x_B,Q^2) ~=~ \int^{Q^2}\frac{dk^2_{\bot}}{k^2_{\bot}} 
\; \vph_g(x_B,k^2_{\bot}) 
~~~~~\mbox{(hereafter} ~q^2=-Q^2,~k^2=-k^2_{\bot} ~~\mbox{)},
\label{1}
 \end{eqnarray}
which satisfies the BFKL equation.

We define  the  Bjorken variables 
\bea
 x_B=Q^2/(2pq)~~~\mbox{ and }~~~x=Q^2/(2kq),
\label{1.1a}
 \end{eqnarray}
for lepton-hadron and lepton-parton scattering, respectively, where 
$p^{\mu}$ and $k^{\mu}$ are the hadron and the gluon 4-momentums, respectively,
and $q^{\mu}$ is the photon 4-momentum.
%$Q^2 =
%- q^2$ is square of photon 4-momentum. 

Notice that the integral is divergent at the lower limit and so it leads to the
necessity to consider the difference $f_{g}(x_B,Q^2) - f_{g}(x_B,Q^2_0)$
with some nonzero $Q^2_0$ (see discussions in Sect. 3), i.e.
 \begin{eqnarray}
f_{g}(x_B,Q^2) ~=~ f_{g}(x_B,Q^2_0) + \int^{Q^2}_{Q^2_0}
\frac{dk^2_{\bot}}{k^2_{\bot}} \; \vph_g(x_B,k^2_{\bot}) 
\label{1.1}
 \end{eqnarray}

In our analysis below we will not use the Sudakov
decomposition, which is sometimes quite convenient in high-energy 
calculations.
However, it is useful to have relations between our calculations and
the results, where the Sudakov decomposition has been used.
%
%The representation of the transverse part $k_{\bot}$of the gluon momentum
%$k$ as a combination of the full momenta $k$ and $q$ will be given
%below (see following section). 
The corresponding analysis will be done in the next Section.
Here we only note that the property
$k^2=-k^2_{\bot}$ (see Eq. (\ref{1})) comes from the fact that
the Bjorken parton variable $x$ in the standard and in the
Sudakov approaches coincide. 
%(where $x=Q^2/2/(kq)$) 

Then, in the BFKL approach the SFs $F^c_{2,L}(x_B,Q^2)$ are driven at small 
$x_B$ by gluons and are related in the following way to the unintegrated 
distribution $\vph_g(x,k^2_{\bot})$: 
\begin{eqnarray}
F^c_{2,L}(x_B,Q^2) ~=~\int^1_{x_B} \frac{dx}{x} \int 
\frac{dk^2_{\bot}}{k^2_{\bot}} 
\; C^g_{2,L}(x,Q^2,m_c^2,k^2_{\bot})~ \vph_g(x_B/x, k^2_{\bot}), 
 \label{d1}
%\nonumber
\end{eqnarray}

The functions $C^g_{2,L}(x,Q^2,m_c^2,k^2_{\bot})$ 
may be regarded as the structure 
functions of
the off-shell gluons with virtuality $k^2_{\bot}$ (hereafter we call them as
{\it coefficient functions}). They are described by the quark box (and
crossed box) diagram contribution to the photon-gluon interaction 
(see Fig. 1). 
\begin{figure}[t]
\begin{center}
\epsfig{figure= 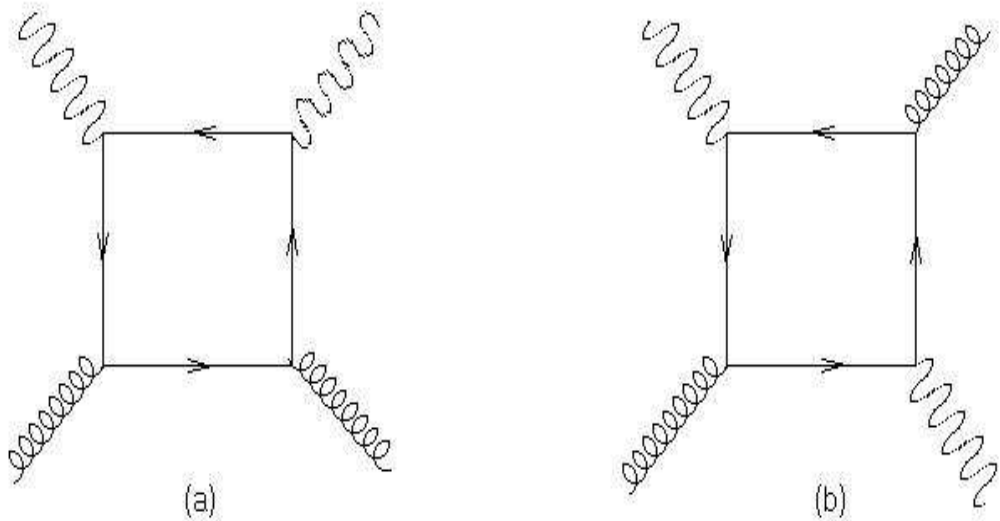,width=16cm,height=12cm}
\end{center}
\caption{ The diagrams contributing to $T_{\mu\nu}$ for a gluon target.
They should be multiplied by a factor of 2 because of the opposite direction 
of the fermion loop. The diagram (a)
%{\bf (a)} 
should be also doubled because of crossing symmetry.}
%The box diagramms of photon gluon fusion.}
\label{fig1}
\end{figure}

The purpose of the article is to calculate these coefficient functions
$C^g_{2,L}(x,Q^2,m_c^2,k^2_{\bot})$ and to analyze experimental data for 
$F^c_{2}(x_B,Q^2)$ by applying Eq. (\ref{d1}) with different sets of
unintegrated gluon densities (see Ref. \cite{LiZo})
and to give predictions for the longitudinal SF $F^c_{L}(x_B,Q^2)$.
%$F_{L}(x_B,Q^2)$.\\

It is instructive to note that
the diagrams shown in Fig. 1. are similar to those  of the
%contributions 
%\footnote{The real difference: in QCD case there is the Casimir factor 
%%$C_F=(N^2-1)/(2N)$ for $SU(N)$ gauge group
%$T_F=n_f/2$, where $n_f$ is the number of active quarks.}
% as in the case of
%Quantum Electrodynamics (QED) in 
photon-photon scattering process.
The corresponding QED contributions have been calculated many years ago
in Ref. \cite{BFaKh} (see also the beautiful review in Ref. \cite{BGMS}). 
Our results 
have been calculated independently 
and they are in full agreement with Ref.
%on 
\cite{BFaKh} (see Appendix B). 
%but the results coincide with them. 
%We would like to note 
However, we hope
that our formulas which are given in a more 
%have essentially 
simple form 
%and we hope that they will 
could be useful for others.\\
%in other analyses.\\

The structure of this article is as follows: in Sect. 2 we
present the basic formalism of our approach with a brief review 
of the calculational steps (based on Ref. \cite{KaKo}). The connection of
our analysis with the
Sudakov-like approach is also given. Later, we present the results for 
the two most important polarization matrices for off-shell gluons into
the proton. 
In Sect. 3 and 4 we give the predictions for the structure functions
$F_2^c$ and $F_L^c$ for two cases of unintegrated gluon
distribution functions (see Ref. \cite{LiZo}) used, which are
shortly reviewed.
 In Appendix A we show the basic technique for the evaluation of the
required Feynman diagrams. Appendix B contains
the review of QED results from Refs. \cite{BFaKh,BGMS}.
In Appendix C
we consider the limiting cases,
when
the values of the quark mass or the gluon momentum are equal to zero 
and also
when the value of the photon ``mass'' $Q^2$ goes to zero.\\

\section{Approach} \indent
The hadron part of the deep inelastic (DIS) spin-average lepton-hadron 
cross section can be represented in the form 
\footnote{Hereafter we consider only one-photon exchange approximation.}:
 \begin{eqnarray}
F_{\mu\nu} ~=~ e_{\mu\nu}(q)~F_{L}(x_B,Q^2) +  d_{\mu\nu}(q,p)
~F_{2}(x_B,Q^2),
 \label{2}
 \end{eqnarray}
where $q^{\mu}$ and $p^{\mu}$ are the photon and hadron momenta,
 \begin{eqnarray}
e_{\mu\nu}(q) = g_{\mu\nu}- \frac{q_{\mu}q_{\nu}}{q^2}~~\mbox{ and }~~
d_{\mu\nu}(q,p) = -\Biggl[ g_{\mu\nu}+ 
2x_B\frac{(p_{\mu}q_{\nu}+p_{\nu}q_{\mu})}{q^2} + 
p_{\mu}p_{\nu}\frac{4x_B^2}{q^2} \Biggr]
\nonumber
\end{eqnarray}
and
$F_k(x_B,Q^2)~~($hereafter $k=2,L)$ are structure functions.

The tensor $F_{\mu\nu}$ is connected via the optical theorem with the 
amplitude of elastic forward scattering of a photon on a hadron
$T_{\mu\nu}(q,p)$, which may be decomposed in invariant amplitudes
$T_k(x_B,Q^2)$ by analogy with Eq. (\ref{2}).

Let us expand the invariant amplitudes in inverse powers of $x_B$:
 \begin{eqnarray}
T_k ~=~ \sum_{n=0}^{\infty} {\biggl(\frac{1}{x_B} \biggr)}^n \; T_{k,n}
%\nonumber
 \label{2a}
 \end{eqnarray}

The coefficients $T_{k,n}$ coincide (for even $n$) with the moments 
$M_{k,n}$ of the SF $F_k$:
 \begin{eqnarray}
T_{k,n} ~=~ M_{k,n} \equiv \int_{0}^{1}dz~ z^{n-2} F_{k}(z,Q^2)~~~ 
%~~~(n=2m)
%\nonumber
 \label{2aa}
\end{eqnarray}

\subsection{Evaluation of coefficient functions} \indent 

We would like to note that the previous formalism can be replicated
at parton level by replacing the hadron momentum $p^{\mu}$ by the
gluon one $k^{\mu}$ and
the Bjorken variable $x_B$ by the corresponding $x$. Then,
the hadron part of the deep inelastic spin-average lepton-parton 
cross section can be represented in the form
 \begin{eqnarray}
F^p_{\mu\nu} ~=~ e_{\mu\nu}(q)~F^p_{L}(x,Q^2) +  d_{\mu\nu}(q,k)
~F^p_{2}(x,Q^2),
 \label{1aa}
 \end{eqnarray}
where 
% \begin{eqnarray}
%e_{\mu\nu} = g_{\mu\nu}- \frac{k_{\mu}k_{\nu}}{k^2}~~\mbox{ and }~~
%d^p_{\mu\nu} = -\Biggl[ g_{\mu\nu}+ 
%2x\frac{(k_{\mu}q_{\nu}+k_{\nu}q_{\mu})}{q^2} + 
%k_{\mu}k_{\nu}\frac{4x^2}{q^2} \Biggr]
%\nonumber
%% \label{1}
% \end{eqnarray}
%and
$F^p_k(x,Q^2)$
%~~($hereafter $k=2,L)$ 
are the structure functions of 
lepton-parton DIS.

As in our analysis we only consider gluons, the unintegrated gluon 
distribution into the parton (i.e. into the gluon) $\vph^p_g(x, k^2_{\bot})$
should have the form
 \begin{eqnarray}
\vph^g_g(x, k^2_{\bot}) ~=~ \delta(1-x) \hat \vph(k^2_{\bot}),
 \nonumber
 \end{eqnarray}
where $\hat \vph(k^2_{\bot})$ is a function of $k^2_{\bot}$.

The parton SF $F^p_k(x,Q^2)$ and the amplitudes $T_k^p(x,Q^2)$ at the parton 
level obey equations similar to
Eq. (\ref{2a}) and Eq. (\ref{2aa}) with the replacement $x_B \to x$. Then, they
are connected via optical theorem:
 \begin{eqnarray}
T_k^p(x,Q^2) ~=~ \sum_{n=0}^{\infty} {\biggl(\frac{1}{x} \biggr)}^n \; 
\int_{0}^{1}dz~ z^{n-2} F_{k}^p(z,Q^2)~~~ 
~~~(n=2m)
%\nonumber
 \label{da}
\end{eqnarray}

Thus, the
coefficient functions $C^g_k(x,Q^2,k^2)$ 
of the parton SF $F^p_k(x,Q^2)$
%(see Eq. (\ref{d1}))
\begin{eqnarray}
F^p_{2,L}(x,Q^2) ~=~
%\int^1_{x} \frac{dy}{y} 
\int \frac{dk^2_{\bot}}{k^2_{\bot}} 
\; C^g_{2,L}(y,Q^2,m_c^2,k^2_{\bot})  \Theta(x_0-x)~ 
\hat \vph(k^2_{\bot}), 
 \label{d1a}
\eea
%$$
%F_k(x,Q^2)= \sum_{i=NS,S,G} \int^1_x \frac{dy}{y} \;  C^i_k(y,Q^2)
%f_i(x/y,Q^2),$$
%where $f_i(x,Q^2)$ are parton distributions, 
can be obtained directly using 
%from calculating the corresponding perturbative parts
%$\tilde C^g_k(x,Q^2,k^2)$ of 
the amplitudes $T_k^p(x,Q^2)$ at the parton level 
\begin{eqnarray}
T^p_{2,L}(x,Q^2) ~=~
%\int^1_{x} \frac{dy}{y} 
\int 
\frac{dk^2_{\bot}}{k^2_{\bot}} 
\; \tilde C^g_{2,L}(y,Q^2,m_c^2,k^2_{\bot})~ 
\hat \vph(k^2_{\bot}), 
 \label{d1aa}
\eea
in the following way:
 \begin{eqnarray}
\tilde C^g_k(x,Q^2,k^2) ~=~  \sum_{n=0}^{\infty} 
{\biggl(\frac{1}{x} \biggr)}^n \; \int_{0}^{1}dz~ z^{n-2} C^g_k(z,Q^2,k^2)
 \Theta(z_0-z),
%\nonumber
 \label{2ab}
 \end{eqnarray}
%Here 
%$\vph_g^p(x,k^2_{\bot})$ is unintegrated gluon distribution in
%considered parton.
where
we have extracted a kinematical factor $\Theta(z_0-z)$.

As it was already discussed we will work with gluon part alone, keeping
nonzero values of  quark masses and the gluon virtuality. The 
corresponding Feynman diagrams are displayed on Fig. 1.
The  coefficient functions $C^g_k(x,Q^2,k^2)$ do not depend on the target
type. So, it can be calculated in photon-parton DIS and used later in 
the photon-hadron reaction (see Eq. (\ref{d1})).\\

\subsection{Connection with the Sudakov-like approach} \indent 

One of the basic ingredients in the Sudakov-like approach is the 
introduction of an additional light-cone momentum $n^{\mu}$ with
$n^2=0$ and $(np)=1$.

The gluon momentum $k^{\mu}$ can be represented as 
\bea
k^{\mu}=\xi p^{\mu} + \frac{k^2 + k^2_{\bot}}{2\xi} n^{\mu} +k_T^{\mu}
\label{2.1}
\eea
with the following properties 
\bea
p^2=n^2=(pk_T)
=(nk_T)=0,~~~(np)=1
%n \cdot p=1
\label{2.2}
\eea
where the four-vector $k^{\mu}_{T}$ contains only the transverse part 
of $k^{\mu}$, $k_T=(0,k_{\bot},0)$, i.e. $k_T^2 = -k^2_{\bot}$.
and $\xi =x_B/x$ is the fraction of the proton momentum carried
by the gluon (see Eq. (\ref{d1}).

To study the relations between the ``usual'' approach used here and
the Sudakov-like one, it is convenient to introduce the following
parametrization for the vector $n^{\mu}$
(see Ref. \cite{EFP}) 
\bea
n^{\mu} ~=~ \frac{2x_B}{Q^2} \left( x_Bp^{\mu} + q^{\mu} \right)
\label{2.3}
 \eea
It is easy to check that the properties in Eq. (\ref{2.2}) are fulfilled.

Then, for the scalar product $(kq)$ we have in the Sudakov-like approach:
\bea
(kq) &=& \xi \; (pq) + \frac{k^2 + k^2_{\bot}}{2\xi} \; (nq) \nonumber \\
&=& 
%\frac{Q^2}{2x}
%\frac{x}{x_B}
%-\frac{(k^2 + k^2_{\bot})x}{2}
%~=~ 
\frac{Q^2}{2x} \left[ 1-
x^2\frac{k^2 + k^2_{\bot}}{Q^2}\right]
\label{2.4}
\eea

If $k^2 = -k^2_{\bot}$, then it follows from Eq. (\ref{2.4})
\bea
%\frac{x_B}{x}
x ~=~ \frac{Q^2}{2(kq)},
%\label{2.5}
\nonumber
\eea
%which is in agreement 
that agrees with Eq. (\ref{1.1a}). Also from Eq. (\ref{2.1}) it follows
\bea
k^{\mu}=xp^{\mu} +k_T^{\mu},
\label{2.1a}
\eea
where $x$ is the fraction of the proton momentum carried by the gluon.\\

\subsection{Feynman-gauge gluon polarization} \indent 

%{\bf 1.} 
As a first approximation  we consider gluons having 
polarization tensor (hereafter
the indices $\alpha$ and $\beta$ are connected with gluons and 
$\mu$ and $\nu$ are connected with photons)\footnote{In 
principle, we can use here more general cases of polarization 
tensor (for example, that one based on the Landau or unitary gauge).
The difference between them and Eq. (\ref{01dd}) is $\sim k^{\alpha}$
and/or $\sim k^{\beta}$
and, hence, it leads to zero contributions because the Feynman diagrams
in Fig.1 are gauge invariant.}: 
\bea
\hat P^{\alpha\beta}=-g^{\alpha\beta}
\label{01dd}
\eea
This polarization tensor corresponds to the case when gluons do not interact.
In some sense the case of polarization is equal to the standard DIS
suggestions about parton properties, excepting their off-shell property. 
The polarization in Eq. (\ref{01dd})
%It 
gives the main
 contribution to the polarization tensor  we are interested in (see below)
\bea
\hat P^{\alpha\beta}_{BFKL}=
\frac{k_{\bot}^{\alpha}k_{\bot}^{\beta}}{k_{\bot}^2}
=\frac{x^2}{-k^2}
p^{\alpha}p^{\beta},~~~ 
%(k=xp+k_t)
\label{1dd}
\eea
which
%It 
comes from the high energy (or $k_T$) factorization prescription
\cite{CaCiHa, CoEllis, LRSS}
\footnote{We would like to note that the BFKL polarization tensor
is a particular case of so-called nonsense polarization of the 
%gauge 
particles in $t$-channel
makes the main contributions for cross sections in $s$-channel at 
$s \to \infty$ (see, for example, Ref. \cite{KurLi} and references therein).
The limit $s \to \infty$ corresponds to the small values of Bjorken variable
$x_B$, that is just the range of our study.}.

Contracting the photon projectors
(connected with photon indices of diagrams
in Fig.1.) 
 \begin{eqnarray}
\hat P^{(1)}_{\mu\nu} = -\frac{1}{2} g_{\mu\nu}~~\mbox{ and }~~
\hat P^{(2)}_{\mu\nu} = 4x^2\frac{k_{\mu}k_{\nu}}{Q^2} 
\nonumber
% \label{1}
 \end{eqnarray}
with the hadronic tensor $F_{\mu\nu}$, we obtain the following relations
at the parton level (i.e. for off-shell gluons having momentum $k_{\mu}$)
 \begin{eqnarray}
\tilde \beta^2 \; C^g_{2}(x) &=& 
%e_c^2 \; \frac{\alpha_s(Q^2)}{4\pi}\; x 
{\cal K}
\;
\left[
f^{(1)} + 
\frac{3}{2\tilde \beta^2}\; f^{(2)} \right]
 \label{3}\\
\tilde \beta^2 \; C^g_{L} (x) &=& 
%e_c^2 \; \frac{\alpha_s(Q^2)}{4\pi}\; x 
{\cal K}
\;
\left[
4bx^2 f^{(1)} + 
\frac{(1+2bx^2)}{\tilde \beta^2}\; f^{(2)} \right]
%\nonumber 
%\label{4}\\
%&=& 
~=~
%e_c^2 \; \frac{\alpha_s(Q^2)}{4\pi}\; x 
{\cal K}
\;
f^{(2)} +4bx^2 \tilde \beta^2\; C^g_2 ,
%= 
%e_c^2 \; \frac{\alpha_s(Q^2)}{4\pi}\; x \; f^{(2)} + 
%\tilde \beta^2 (1-\tilde \beta^2) \; C^g_2, 
 \label{4}
%\nonumber
 \end{eqnarray}
where the normalization factor 
${\cal K} =  e_c^2 \; \alpha_s(Q^2)/(4\pi)\; x$, 
$$P^{(i)}_{\mu\nu} F_{\mu\nu} =
%e_c^2 \; \frac{\alpha_s(Q^2)}{4\pi}\; x 
{\cal K} \;  f^{(i)} 
, \, i = 1, 2$$ and 
$\tilde \beta^2=1-4bx^2,~~b=-k^2/Q^2 \equiv  k_{\bot}^2/Q^2 >0,~~
a=m^2/Q^2$.
The kinematical factor $z_0$ which appear in  Eq. (\ref{2ab}) is
 \begin{eqnarray}
z_0~=~\frac{1}{1+4a+b}
%\nonumber
 \label{1A}
 \end{eqnarray}

Applying the projectors $\hat P^{(i)}_{\mu\nu}$ to the Feynman diagrams  
displayed in Fig.1, we obtain
\footnote{ The contributions of individual scalar 
components of the diagrams of Fig.1 
(which come after evaluation of traces of $\gamma$-matrices)
are given in Appendix A.} 
the following
results for the contributions to expressions 
%$$F^{(i)}=e_c^2x \cdot \frac{\alpha_s(Q^2)}{4\pi}\cdot f^{(i)}$$
%and
\begin{eqnarray}
f^{(1)} &=& -2 \beta \Biggl[ 1 - \biggl(1-2x(1+b-2a) \; [1-x(1+b+2a)] 
\biggr) \; f_1  
\nonumber \\
    &+& (2a-b)(1-2a)x^2 \; f_2  \Biggr],
 \label{5}\\
f^{(2)} &=& 8x\; \beta \Biggl[(1-(1+b)x)  
-2x \biggl(bx(1-(1+b)x)(1+b-2a) + a\tilde \beta^2 \biggr)\; f_1  
\nonumber \\
    &+& bx^2(1-(1+b)x) (2a-b) \; f_2  \Biggr],
\label{6} 
\end{eqnarray}
where
$$
%b=\frac{p^2}{Q^2},~~
%~~\tilde \beta^2=1+4bx^2,
%~~~~~~
\beta^2=1-\frac{4ax}{(1-(1+b)x)}$$
and \footnote{ We use the variables as defined in Ref. \cite{Vog}.}
$$f_1=\frac{1}{\tilde \beta \beta} \;
ln\frac{1+\beta \tilde \beta}{1-\beta \tilde \beta},
~~~~~f_2=\frac{-4}{1-\beta^2 \tilde \beta^2}$$ 
\vskip 0.5cm

The important regimes: 
$k^2 = 0$, $m^2 = 0$ and $Q^2 \to 0$
are considered  in Appendix C.
The $Q^2 = 0$ limit is given in Sect. 2.5.\\

\vskip 0.5cm

\subsection{BFKL-like gluon polarization} \indent 

%{\bf 2.}
Now we take into account 
the BFKL gluon polarization given in Eq. (\ref{1dd}). As we 
already noted in previous subsection, in these calculations we did not use
Sudakov decomposition and, hence, the hadron momentum $p^{\alpha}$ is not so
convenient variable in our case. Thus, we  represent 
the projector $\hat P^{\alpha\beta}_{BFKL}$
as a combination of projectors constructed by the momenta $k^{\alpha}$ and 
$q^{\alpha}$.

We can represent the tensor $F^{\alpha \beta}$ in the general form: 
\bea
F^{\alpha \beta} ~=~Ag^{\alpha \beta}+Bq^{\alpha}q^{\beta} 
+Ck^{\alpha}k^{\beta} + 
D \left(k^{\alpha}q^{\beta} +q^{\alpha}k^{\beta} \right),
\label{7.1} 
\eea
where
$A$, $B$, $C$ and $D$ are some scalar functions of the variables $y$, $a$ 
and $b$. 

>From the gauge invariance of the vector current: 
$k^{\alpha}F^{\alpha \beta}=k^{\beta}F^{\alpha \beta}=0$ we have the 
following relations
\bea
C \; k^2 ~=~-\left[A + D \; (kq) \right],~~~
B\; (kq) ~=~- D \; k^2
\label{7.2} 
\eea

If we apply the BFKL-like projector $\hat P^{\alpha\beta}_{BFKL}$
and use the light-cone properties given in Eq. (\ref{2.2}), we get
the simple relation 
\bea
\hat P^{\alpha\beta}_{BFKL} F^{\alpha \beta} ~=~ D \; (kq)
\label{7.3} 
\eea

The standard projectors $g^{\alpha \beta}$ and $q^{\alpha}q^{\beta}$
lead to the relations
\bea
g^{\alpha \beta} F^{\alpha \beta} &=& 3A + D \frac{Q^2}{2x}\tilde \beta^2
\label{7.4} \\
q^{\alpha}q^{\beta}F^{\alpha \beta} &=& \frac{\tilde \beta^2}{bx^2} 
\left[ 3A + D \frac{Q^2}{2x}\tilde \beta^2 \right]
\label{7.5}
\eea

>From Eqs. (\ref{7.2}), (\ref{7.4}) and (\ref{7.5}) we have
\bea \left[ \left( (kq)^2 -k^2q^2 \right) g^{\alpha \beta} 
+3 k^2 q^{\alpha}q^{\beta} \right]
F^{\alpha \beta} ~=~ - D \frac{Q^8}{4x^4}\tilde \beta^4
\label{7.6}
\eea
and the BFKL-like projector $\hat P^{\alpha\beta}_{BFKL}$
can be represented as
\bea
\hat P^{\alpha\beta}_{BFKL} ~=~ -\frac{1}{2}
\frac{1}{\tilde \beta^4} \left[\tilde \beta^2 g^{\alpha \beta} 
-12 bx^2 \frac{q^{\alpha}q^{\beta}}{Q^2} \right]
\label{3dd} 
\eea

%We
%note that the relation (\ref{3dd}) is a property of boson-gluon fusion.
%Indeed, to obtain (\ref{7.2}) we use the gauge invariance of vector 
%current.

%Because (\ref{2dd}), we can represent the projector 
%$\hat P^{\alpha\beta}_{BFKL}$ in the following form:
%\bea
%\hat P^{\alpha\beta}_{BFKL}=\frac{1}{\tilde\beta^4}
%\left[
%\tilde\beta^2 \cdot 
%\left(-frac{1}{2} g^{\alpha\beta} \right) ~+~
%3bx^2 q^{\alpha}q^{\beta} 
%\right]
%\label{3dd}
%\eea

In the previous section we have already calculated the contributions to
coefficient functions using the first term within the brackets in
the r.h.s. of Eq. (\ref{3dd}).
Repeating the above calculations with the projector
$\sim q^{\alpha}q^{\beta}$, we obtain the total contribution to
the coefficient functions which can be represented as the following
shift in the results given in Eqs. (\ref{3})-
%,\ref{4},\ref{5}) and 
(\ref{6}):
\bea
C^g_{2}(x)~ &\to& ~ C^g_{2,BFKL}(x),~~~C^g_{L}(x)~ \to ~ C^g_{L,BFKL}(x);
\nonumber \\
 f^{(1)}~ &\to& ~f^{(1)}_{BFKL}= 
\frac{1}{\tilde\beta^4}
\left[ \tilde \beta^2 f^{(1)}  ~-~
3bx^2 \tilde f^{(1)}\right] \nonumber \\
 f^{(2)} ~&\to&~ f^{(2)}_{BFKL}=
\frac{1}{\tilde\beta^4}
\left[ \tilde \beta^2 f^{(2)}  ~-~
3bx^2 \tilde f^{(2)}
\right],
\label{4dd}
\eea
where
\begin{eqnarray}
\tilde f^{(1)} &=& - \beta \Biggl[ \frac{1-x(1+b)}{x}  
-2 \biggl(x(1-x(1+b))(1+b-2a) +a \tilde \beta^2 \biggr) \; f_1  
\nonumber \\
    &-& x(1-x(1+b))(1-2a) \; f_2  \Biggr],
 \label{5dd}\\
\tilde f^{(2)} &=& 4 \; \beta ~(1-(1+b)x)^2 \Biggl[2   
- (1+2bx^2)\; f_1  
%\nonumber \\    &-& 
-bx^2 \; f_2  \Biggr],
\label{6dd} 
\end{eqnarray}

For the important regimes when
$k^2= 0$, $m^2 = 0$ and $Q^2 \to 0$, the analyses are given in 
Appendix C.

Notice that our results in Eqs. (\ref{5dd}) and (\ref{6dd}) should
coincide with the
integral representations of Refs. \cite{CaCiHa, CaCiHa2} (at $Q^2 \to 0$ 
there is full agreement (see following subsection)
with the formulae of Refs. \cite{CaCiHa, CaCiHa2} for photoproduction of heavy
quarks). Our results in Eqs. (\ref{5dd}) and (\ref{6dd}) should also
agree with those in Ref. \cite{BMSS} but the direct comparison is 
quite difficult because the
authors of Ref. \cite{BMSS} used a different (and quite complicated) way to 
obtain their results and the
structure of their results is quite cumbersome (see Appendix A in Ref. 
\cite{BMSS}). 
%\\
We have found numerical agreement in the case of 
$F_2(x_B,Q^2)$ (see Sect. 3 and Fig.4).\\

\vskip 0.5cm

\subsection{$Q^2 = 0$ limit and Catani-Ciafaloni-Hautmann approach} \indent 

We introduce the new variables $\hat s$, $\rho$ and $\Delta$ which are
useful in the limit $Q^2 \to 0$:
\bea
\hat s=\frac{Q^2}{x}, ~~ \rho=
%\equiv 
4ax \equiv 
%\frac{4m^2}{Q^2}x=
\frac{4m^2}{\hat s}, ~~ 
\Delta =
%\equiv 
bx \equiv \frac{-k^2}{Q^2}x=\frac{k^2_{\bot}}{\hat s}
 \label{c1}
\eea
and express our formulae above as functions of $\rho$ and $\Delta$ at
small $x$ asymptotic (i.e. small $Q^2$).

%{\bf a)} 
When $x = 0$ we have got the following relations:\\
for the intermediate functions
\bea 
\tilde \beta^2 &=& 1,~~ \beta^2 = 1-\frac{\rho}{1-\Delta} 
\equiv \hat \beta^2, ~~
f_1=\frac{1}{\hat \beta }ln\frac{1+ \hat \beta }{1-\hat \beta } 
\equiv L(\hat \beta),~~
f_2=\frac{4\hat \beta }{\rho}(1-\Delta)
 \label{c2.1}\\
& & \nonumber \\
f^{(1)} &=& 2 \hat \beta \Biggl[ \left(1+\rho -\frac{\rho^2}{2}\right) 
L(\hat \beta)
-(1-\rho)
%\nonumber \\&+& 
+
\left(2+\rho -2 L(\hat \beta) \right)\Delta
\nonumber \\&+&
2 \left(L(\hat \beta)-1 \right)\Delta^2 \Biggr],
 \label{c2.2}\\
x\tilde f^{(1)} &=& - 2\hat \beta \Biggl[ 2(1-\Delta)- \rho L( \hat \beta)
\Biggr],~~~
%\nonumber \\
f^{(2)} ~=~
%&=& 0, ~~ 
x \tilde f^{(2)}~=~0, 
 \label{c2.3}
\eea
and, thus, for  the coefficient functions
\bea
C^g_{L} ~=~ 0, ~~~~~ 
%\label{c3.1}\\
C^g_{2} ~=~ 
%e_c^2 \; \frac{\alpha_s(Q^2)}{4\pi}\cdot x 
{\cal K} \;
f^{(1)}_{BFKL},
 \label{c3.2}
 \end{eqnarray}
where
\bea 
f^{(1)}_{BFKL} &=& 2 \hat \beta \Biggl[ 
\left(1+\rho -\frac{\rho^2}{2}\right) L(\hat \beta)
-(1+\rho)
%\nonumber \\ &+& 
+
\left(8+\rho -(2+3\rho) L(\hat \beta) \right)\Delta
\nonumber \\ &+& 
2 \left(L(\hat \beta)-4 \right)\Delta^2 \Biggr],
 \label{c3.3}
\eea
 
We note that the results coincide exactly with those from
Catani-Ciafaloni-Hautmann 
%(CCH)
work in Ref. \cite{CaCiHa, CaCiHa2} (see Eq. (2.2) in Ref. \cite{CaCiHa2})
in the case of photoproduction of heavy quarks.
The $O(x)$ contribution in the $Q^2 \to 0$ limit is given in Appendix C (see
subsection C.3).\\

\section{Comparison with $F_2^c$ experimental data } \indent

%\vskip 0.5cm
With the help of the results obtained in the previous Section
we have analyzed HERA data for SF $F_2^c$ from ZEUS \cite{ZEUS2} and
H1 \cite{H1c} collaborations.\\

\subsection{Unintegrated gluon distribution} \indent 

In this paper we consider two different parametrizations for the
unintegrated gluon distribution ~\cite{LiZo}. Firstly, 
we use the parametrization based on the numerical solution of the BFKL
evolution equation~\cite{RS} (RS--parametrization).
The solution has the following form~\cite{RS}:
\begin{eqnarray}
\Phi(x, k^2)&{=}& \frac{a_1}{a_2 + a_3 +a_4}
                  \left[a_2 + a_3 \;  \frac{Q_0^2}{k^2} 
                + \left(\frac{Q_0^2}{k^2}\right)^2 
%\nonumber \\
%&& 
             + \alpha \; x + \frac{\beta}{\epsilon + \ln (1/x)}\right]
\nonumber \\
&&                 C_q{\left[\frac{a_5}{a_5 + x}\right]}^{1/2}
% \nonumber \\
%&&              
\left[1 -  a_6 x^{a_7}\ln{(k^2/a_8)}\right](1 + a_{11} x) 
%\nonumber \\
%&&  
            (1 - x)^{a_9 + a_{10}\ln (k^2/a_8)},
\label{zo1}
\end{eqnarray}
where
\bea
C_q = \cases{1, & $\mbox{if} \quad k^2  < q_0^2(x)$, \cr
  q_0^2(x)/k^2, & $\mbox{if} \quad k^2  > q_0^2(x)$}
\label{zo1d}
\eea
The parameters $(a_1 - a_{11}, \alpha, \beta$ and
$\epsilon)$ were found (see~Ref. \cite{RS}) by minimization
of the differences between
the l.h.s and the r.h.s. of the BFKL-type equation for the 
unintegrated gluon distribution $\Phi (x, k^2)$ with `$Q_0^2 =$4 GeV$^2$.

Secondly, we also use the results of a BFKL-like parametrization
of the unintegrated gluon distribution 
$\Phi(x, k_{\bot}^2, \mu^2)$, according to the prescription given 
in~Ref. \cite{BL}. 
The proposed method lies upon a straightforward perturbative solution of
the
BFKL equation where the collinear gluon density $x\,G(x,\mu^2)$
from the standard GRV set~\cite{GRV} is used as
the boundary condition in the integral form of Eq. (1).
Technically, the unintegrated gluon density is calculated as a convolution
of the collinear gluon density $G(x,\mu ^2)$
with universal weight factors~\cite{BL}:
\bea
 \Phi(x, k_{\bot}^2, \mu^2) = \int_x^1
 {\cal G}(\eta, k_{\bot}^2, \mu^2)\,
 \frac{x}{\eta}\,G(\frac{x}{\eta},\mu^2)\,d\eta,
\label{zo1dd}
\eea
where
%\begin{equation} 
% {\cal G}(\eta,  k_{\bot}^2, \mu^2)=\frac{\bar{\alpha}_s}{\eta k_{\bot}}\,
% J_0(2\sqrt{\bar{\alpha}_s\ln(1/\eta)\ln(\mu^2/k_{\bot})}),  \\
%\end{equation}
% if $k_{\bot} < \mu^2$, 
%
%\begin{equation}
% {\cal G}(\eta,  k_{\bot}^, \mu^2)=\frac{\bar{\alpha}_s}{\eta k_{\bot}}\,
% I_0(2\sqrt{\bar{\alpha}_s\ln(1/\eta)\ln(k_{\bot}/\mu^2)}),  \\
%\end{equation}
% if $k_{\bot} > \mu^2$, 
\bea
 {\cal G}(\eta,  k_{\bot}^2, \mu^2)=\frac{\bar{\alpha}_s}{\eta k_{\bot}}\,
 \cases{ J_0(2\sqrt{\bar{\alpha}_s\ln(1/\eta)\ln(\mu^2/k^2_{\bot})}), 
& $\mbox{if} \quad k^2_{\bot}  < \mu^2$, \cr
 I_0(2\sqrt{\bar{\alpha}_s\ln(1/\eta)\ln(k^2_{\bot}/\mu^2)})
, & $\mbox{if} \quad k^2_{\bot}  > \mu^2$},
\label{zo2}
\eea20xun01
where $J_0$ and $I_0$ stand for Bessel functions (of real and imaginary
arguments, respectively), and $\bar{\alpha}_s=3{\alpha}_s/\pi$.
The  parameter $\bar{\alpha}_s$ is connected with
 the Pomeron trajectory intercept:
$\Delta_P=\bar{\alpha}_s4\ln{2}$ in the LO and 
$\Delta_P=\bar{\alpha}_s4\ln{2}-N\bar{\alpha}_s^2$ in the NLO approximations,
where $N$ is a number, $N \sim 18$  \cite{FaLi}-\cite{Ross}. 
However, some resummation procedures proposed in the last years
lead to positive value of $\Delta_P \sim 0.2 - 0.3$
%~\cite{Salam},\cite{BFKLP}.
(see Refs. \cite{Salam,BFKLP} and references therein).

Therefore, in our calculations with Eq. (\ref{zo1dd}) we only used
the solution of the LO BFKL equation and 
considered $\Delta_P$ as a free parameter varying it from 0.166 to 0.53.
This approach was used for the description of the $p_T$ spectrum
of $D^*$ meson electroproduction at HERA~\cite{BaZo} where the value
for the Pomeron intercept parameter $\Delta_P =0.35$ was obtained
\footnote{ Close values for the parameter 
 $\Delta_P$ were obtained, rather,
in very different papers (see, for example, Ref. \cite{KMS})
 and in the L3 experiment \cite{L3}.}.
We used this value of  $\Delta_P$ in our present calculations
with $\mu^2 = Q_0^2 = 1-4$ GeV$^2$.

%\vskip 1.5cm

\subsection{Numerical results} \indent
\begin{figure}
\begin{center}
\epsfig{figure=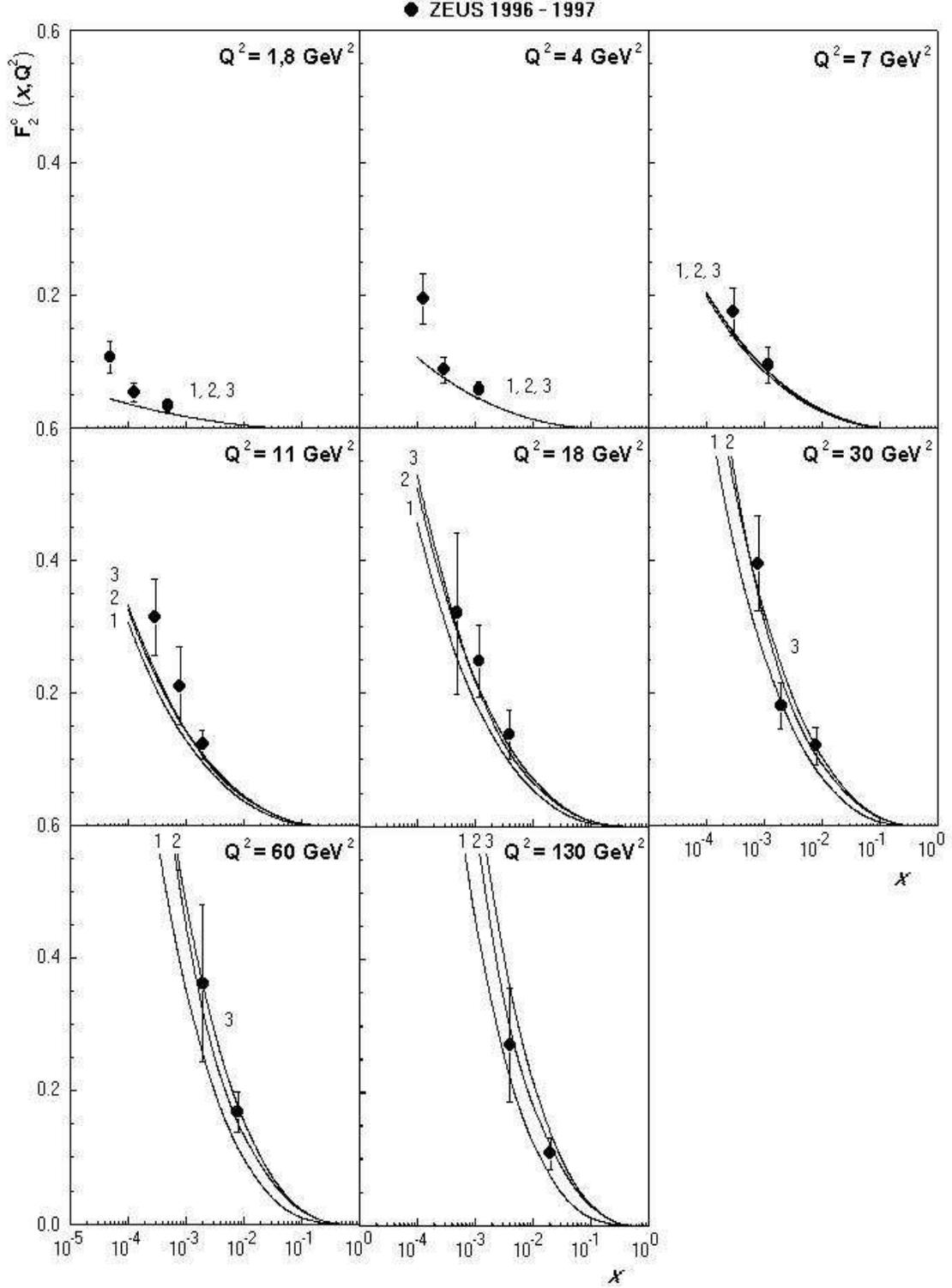,width=16cm,height=21.2cm}
\end{center}
\vskip -1cm
\caption{The structure function $F_2^c(x,Q^2)$ as a function of $x$ for
different values of $Q^2$ compared to ZEUS data~\cite{ZEUS2}.
Curves 1, 2 and 3 correspond to SF obtained
in the standard parton model with the GRV ~\cite{GRV}
gluon density at the leading order approximation and to SF
obtained in the $k_T$ factorization approach with RS~\cite{RS} 
and BFKL (at $Q_0^2 = 4$ GeV$^2$)~\cite{BL} parametrizations of 
unintegrated gluon distribution.}
\label{fig2}
\end{figure}
For the calculation of the SF $F_2^c$ we use Eq. (\ref{d1})
in the  following form:
\bea
F_2^c(x, Q^2)
% = \sum_{i=1}^3 F_2^{c(i)}(x, Q^2), 
%\label{zo3}
%\eea  
%where
%\begin{eqnarray}
%F_2^{c(i)}(x, Q^2) 
&=&\int_{x(1+4a)}^{1} \frac{dy}{y}
C_2^g(\frac{x}{y}, Q^2, 0) \, yG(y, Q_0^2) +  \nonumber \\
&+& \sum_{i=1}^2
\int_{y_{min}^{(i)}}^{y_{max}^{(i)}} \frac{dy}{y} \int_
{k_{{\bot}min}^{2(i)}}^{k_{{\bot}max}^{2(i)}} dk_{\bot}^2 \,
C_2^g(\frac{x}{y}, Q^2, k_{\bot}^2) \, \Phi (y, k_{\bot}^2, Q_0^2). 
\label{zo4}
\end{eqnarray}
Here $\Phi (y, k_{\bot}^2, Q_0^2) = \frac{1}{k_{\bot}^2} \varphi_{g}
(y, k_{\bot}^2, Q_0^2)$ and the changes $x_{B} \to x, x \to y$ were
done in comparison with Eq. (\ref{d1}). The coefficient function $C_2^g(x)$ 
is given in Eq. (\ref{3}) with $f^{(1)}_{BFKL}$ and $f^{(2)}_{BFKL}$ instead 
of $f^{(1)}$ and
 $f^{(2)}$, respectively (see Eq. (\ref{4dd})). 
The functions $f^{(1)}$ and $f^{(2)}$ in Eq. (\ref{3})
are given by Eqs.  (\ref{5}) and (\ref{6}), respectively, and  
the functions $\tilde f^{(1)}$ and $\tilde f^{(2)}$ can be found in
Eqs. (\ref{5dd}) and (\ref{6dd}), respectively.
\begin{figure}[t]
\begin{center}
\epsfig{figure=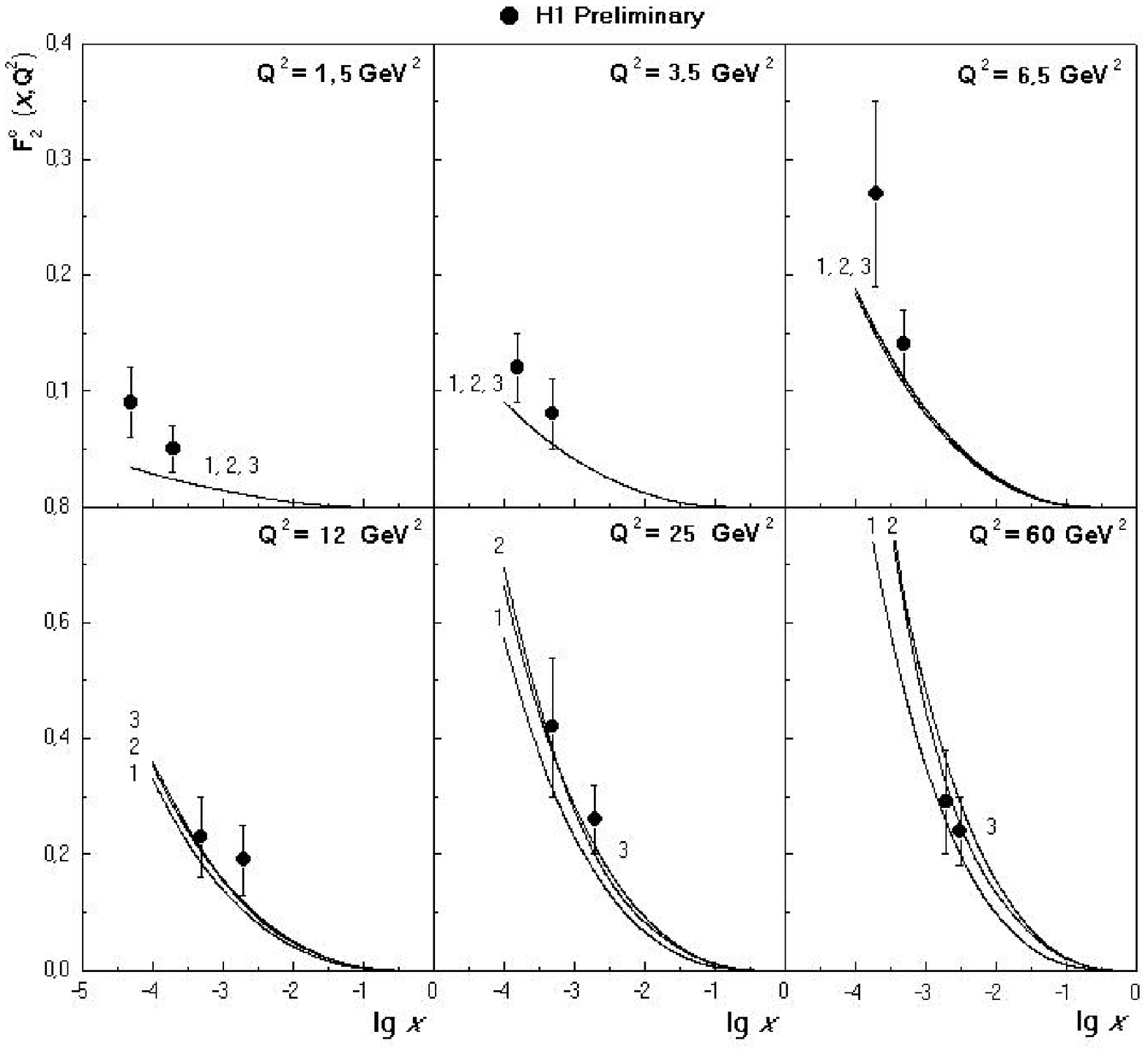,width=16cm,height=12cm}
\end{center}
\caption{The structure function $F_2^c(x,Q^2)$ as a function of $x$ for
different values of $Q^2$ compared to H1 data~\cite{H1c}.
Curves 1, 2 and 3 are as in Fig. 2.}
\label{fig3}
\end{figure}

The integration limits in Eq. (\ref{zo4}) have the following
values:
\begin{eqnarray}
y^{(1)}_{min} &=& x(1 + 4a + \frac{Q_0^2}{Q^2}), \,\,\;\;  y^{(1)}_{max} ~=~
2x(1 + 2a);  \nonumber \\
k_{{\bot}min}^{2(1)} &=& Q_0^2, \,\,\;\; k_{{\bot}max}^{2(1)} ~=~
(\frac{y}{x} - (1 +4a))Q^2; \nonumber \\
y^{(2)}_{min} &=& 2x(1 + 2a), \,\,\;\; y^{(2)}_{max} ~=~ 1; \nonumber \\
k_{{\bot}min}^{2(2)} &=& Q_0^2, \,\,\;\; k_{{\bot}max}^{2(2)} ~=~ Q^2; 
\end{eqnarray}

\begin{figure}[t]
\begin{center}
\epsfig{figure=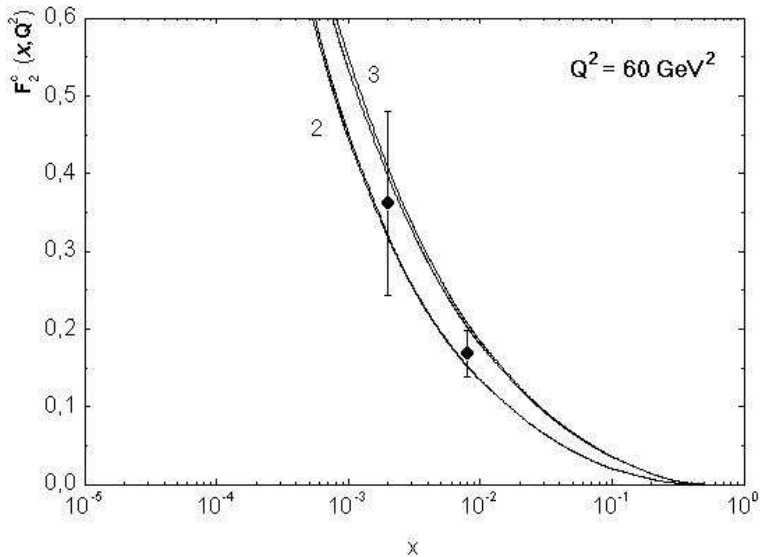,width=11cm,height=9cm}
\end{center}
\caption{The structure function $F_2^c(x,Q^2)$ as a function of $x$ 
at  $Q^2 = 60$ GeV$^2$ compared to ZEUS data~\cite{ZEUS2}. Curves 2 and 3
correspond to RS (at $Q_0^2 = 4$ GeV$^2$)~\cite{RS} and 
BFKL (at $Q_0^2 = 1$ GeV$^2$)~\cite{BL}
parametrizations obtained with our off mass shell matrix  and
ones from Ref.~\cite{BMSS}.}
\label{fig4}
\end{figure}

The ranges of integration correspond to the requirement of
positive values in the arguments of the
square roots in Eqs.  (\ref{5}), (\ref{6}), (\ref{5dd}) and (\ref{6dd})
and also obey to the kinematical restriction $(z \equiv x/y) \leq z_0$ 
with $z_0$ from Eq. (\ref{1A}).

\begin{figure}
\begin{center}
\epsfig{figure=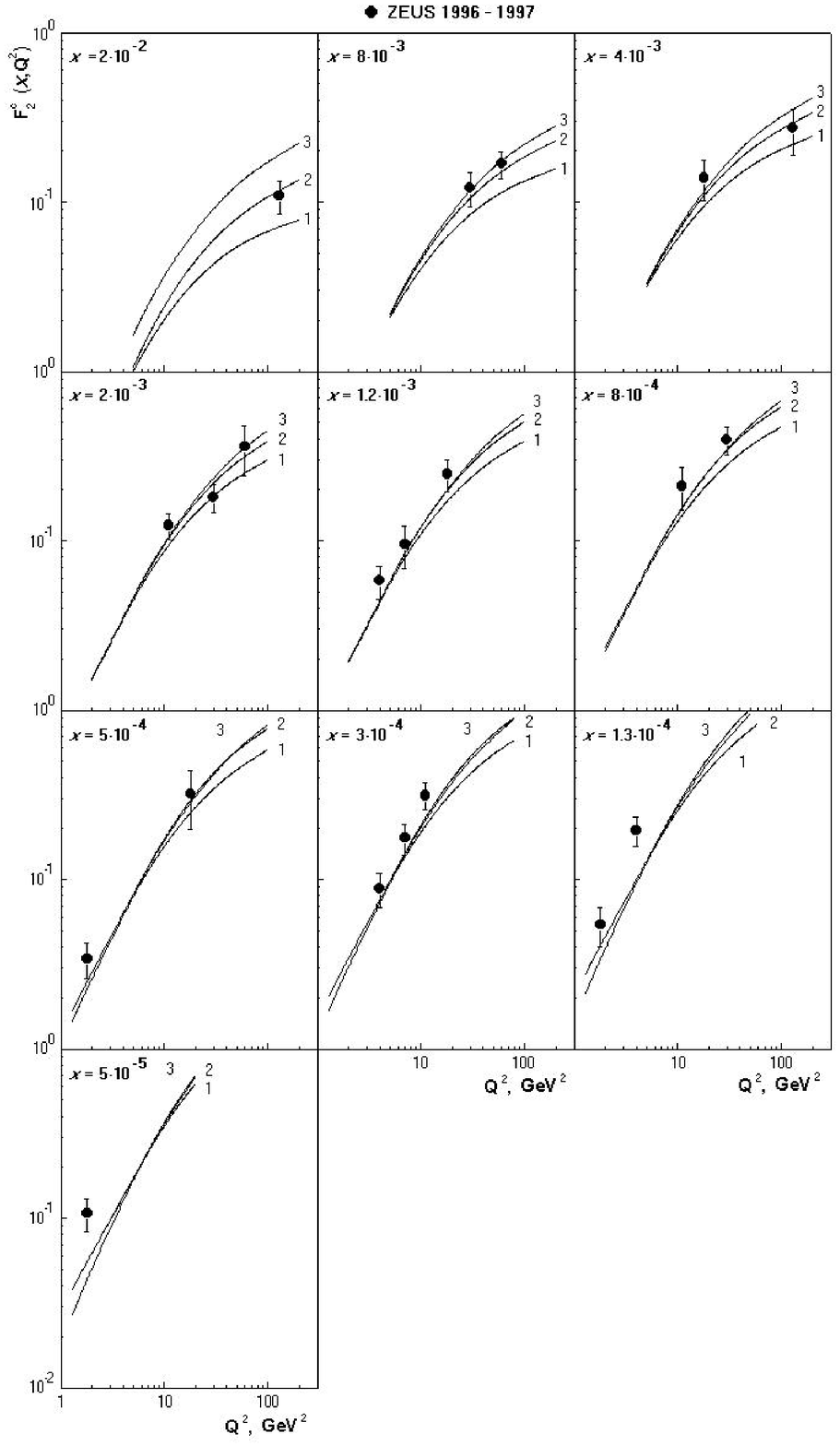,width=16cm,height=22cm}
\end{center}
\vskip -0.5cm 
\caption{The structure function $F_2^c(x,Q^2)$ as a function of $Q^2$ 
for different values of $x$ compared to ZEUS data~\cite{ZEUS2}.
Curves 1, 2 and 3 are as in Fig. 2.}
\label{fig5}
\end{figure}

In Figs. 2 and 3 we show  the SF $F_2^c$ as a function $x$ for different values
of $Q^2$  in comparison with ZEUS \cite{ZEUS2} and H1 \cite{H1c}
experimental data.
For comparison we present the results of the calculation with two different 
parametrizations for the unintegrated gluon distribution $\Phi (x,
k^2_{\bot}, Q_0^2)$ in the forms given by Eq. (\ref{zo1})\, and Eq.
(\ref{zo1dd}) at $Q_0^2 = 4$ GeV$^2$.

 The differences observed between the curves 2 and 3 are 
 due to the different behaviour of the unintegrated gluon distribution
as function $x$ and $k_{\bot}$.

\begin{figure}[t]
\begin{center}
\epsfig{figure=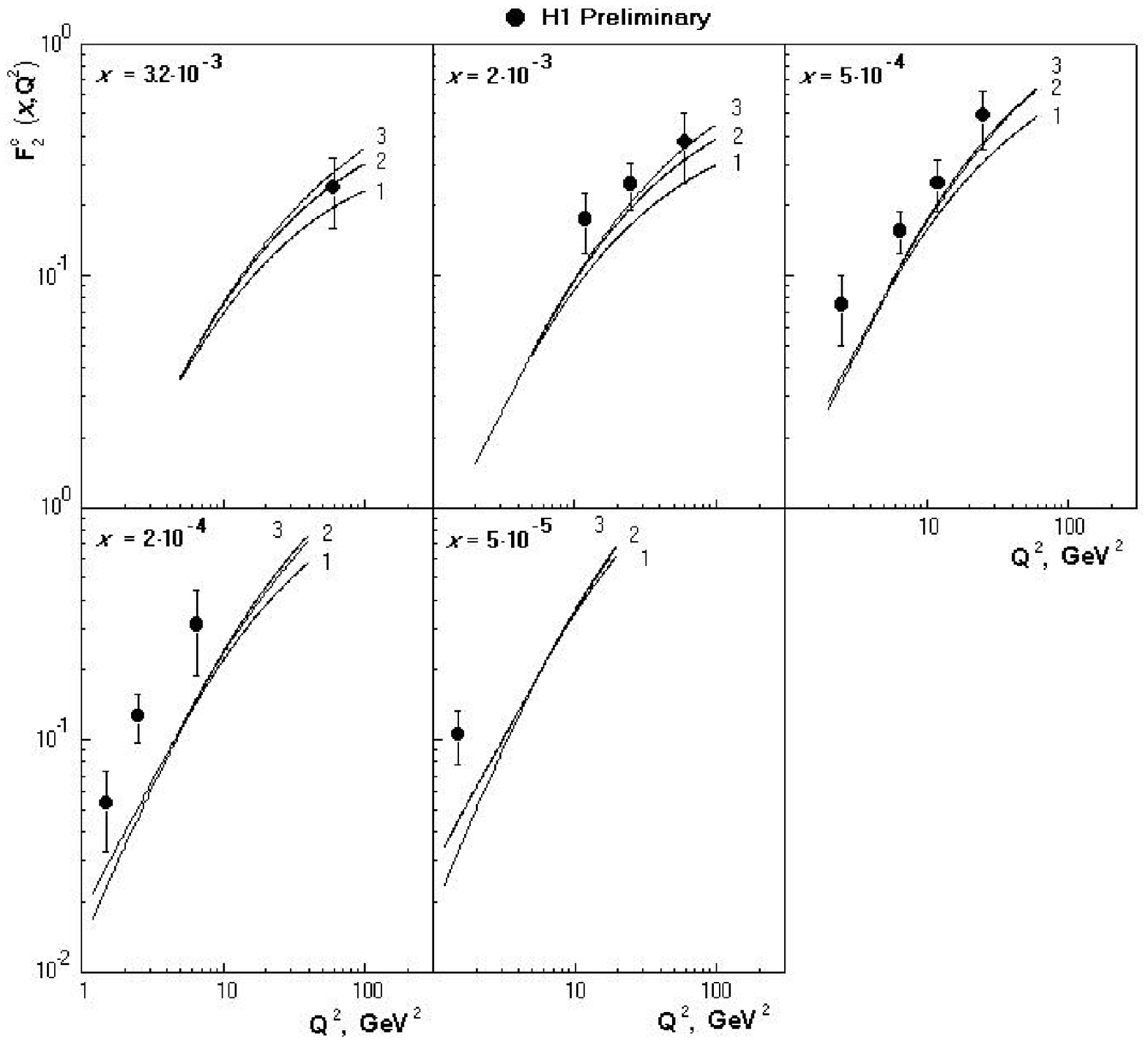,width=16cm,height=13cm}
\end{center}
\caption{The structure function $F_2^c(x,Q^2)$ as a function of $Q^2$ 
for different values of $x$ compared to H1 data~\cite{H1c}.
Curves 1, 2 and 3 are as in Fig. 2.}
\label{fig6}
\end{figure}

 We see that at large $Q^2$ ($Q^2 \geq 10$ GeV$^2$)
the SF $F_2^c$
obtained in the $k_T$ factorization approach is higher than the SF obtained
in the standard parton model with the GRV 
%and MT ~\cite{MT} 
gluon density at the LO approximation 
(see curve 1) 
and has a more rapid growth in comparison with the standard parton model
results, especially at $Q^2 \sim 130$ GeV$^2$~\cite{LZ}.
At $Q^2 \leq 10$ GeV$^2$ the predictions from
perturbative QCD (in GRV approach)
and those based on the $k_T$ factorization approach are very similar
\footnote{This fact is due to the quite large value of $Q^2_0=4$ 
GeV$^2$ chosen here.} and show the disagreement with data 
below $Q^2 = 7$ GeV$^2$
\footnote{A similar disagreement with data at $Q^2 \leq 2$ GeV$^2$
has been observed for the complete structure function $F_2$
(see, for example, the discussion in Ref. \cite{Q2evo}
and reference therein). 
We note that the insertion of higher-twist corrections in the framework
of usual perturbative QCD improves the agreement with data
(see Ref. \cite{HT}) at quite low values of $Q^2$.}.
Unfortunately the available experimental data do not permit yet
to distinguish 
%so far 
the $k_T$ factorization effects from
those due to boundary conditions~\cite{RS}.

Fig. 4 shows the structure function 
%a comparison between  our SF  
$F_2^c$ at $Q^2=60$ GeV$^2$
%with the SF 
obtained with two
different gluon densities, i.e. RS  (at $Q_0^2 = 4$ GeV$^2$)
and BFKL (at $Q_0^2 = 1$ GeV$^2$) parametrizations.
The difference between curves 2 and 3 are mainly due to the different
$Q_0^2$ value used (as we have already shown
in Figs. 2 and 3, the difference due to the
parametrizations is essentially smaller).
From Figs. 2, 3 and 4,
we note that the difference between the $k_T$ factorization results and
those from perturbative QCD increases when we change the value of
$Q_0^2$ in Eq. (3) from 4 GeV$^2$  to 1 GeV$^2$~\cite{LZ}.
In addition, for each case presented in Fig. 4. we have done the 
calculations with our off mass shell matrix elements and those
from Ref.~\cite{BMSS}
\footnote{We would like to note that Ref. \cite{BMSS} contains several
slips: the propagators in Eq. (A.1) and the products $(pq)$ in
Eqs. (A.4) and (A.5) should be in the denominator,
the indices 2 and $L$ in Eqs. (A.4) and (A.5) should be transposed.}.
The predictions are very similar and cannot be distinguished on curves 2
and 3.

For completeness, in Figs. 5 and 6 we present the SF $F_2^c$ as a
function $Q^2$ for different values of $x$ in comparison with ZEUS \cite{ZEUS2}
and H1 \cite{H1c} experimental data.

\vskip 0.5cm 
\section{Predictions for $F_L^c$ } \indent
\begin{figure}
%\vskip -2.5cm 
%\vspace{-2.5cm}
\begin{center}
\epsfig{figure=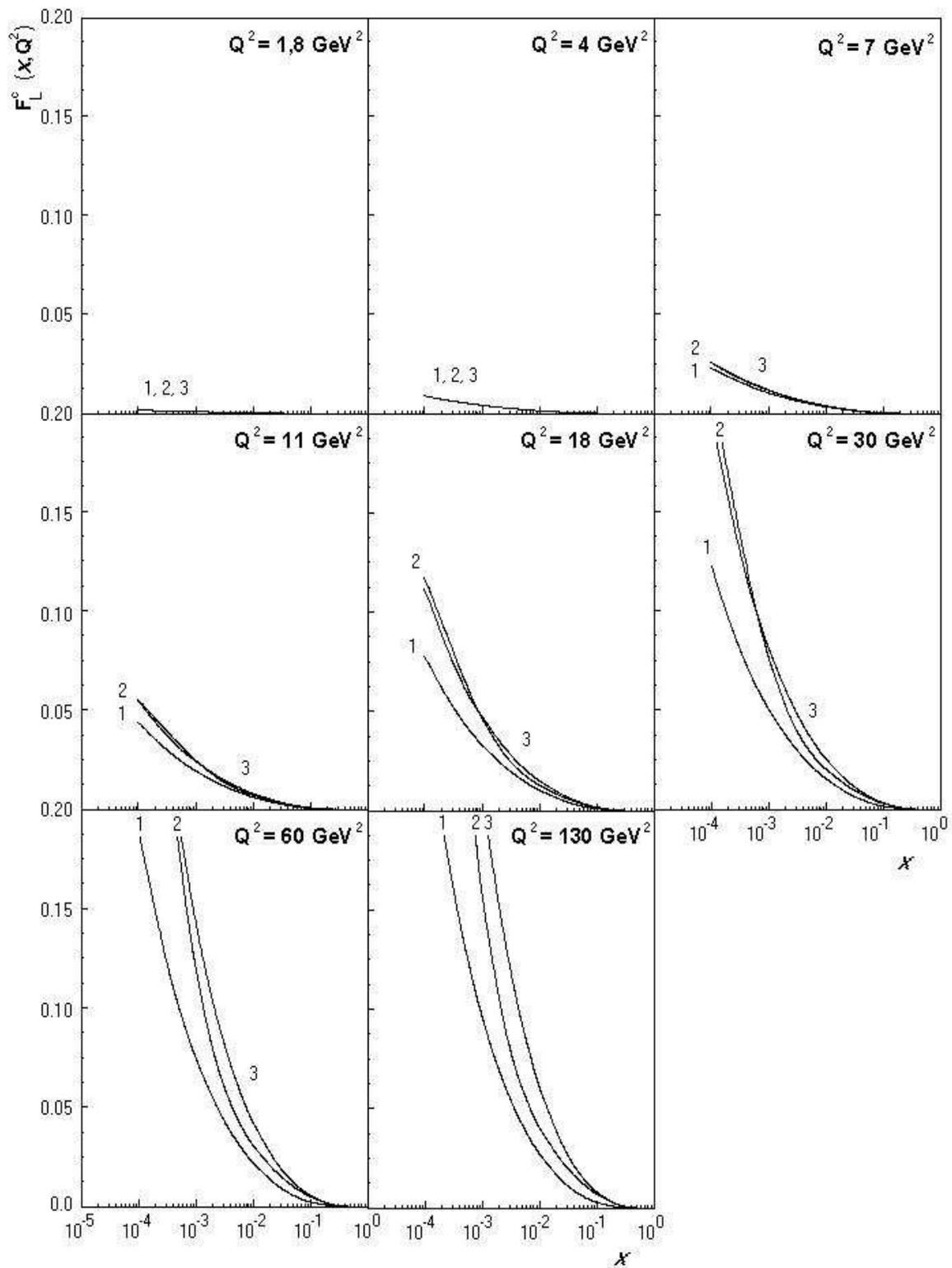,width=16.5cm,height=22cm}
\end{center}
\vskip -0.5cm 
\caption{The structure function $F_L^c(x,Q^2)$ as a function of $x$ 
for different values of $Q^2$.
Curves 1, 2 and 3 are as in Fig. 2.}
\label{fig7}
\end{figure}

To calculate the SF  $F_L^c$ we have used 
Eq. (\ref{zo4})
with the replacement of the coefficient function $C_2^g$ by  $C_L^g$, which
is given by Eq. (\ref{4}).

In Fig. 7 we show the predictions for  $F_L^c$ obtained with different 
unintegrated gluon distributions. 
The difference between the results obtained in perturbative QCD and 
from the $k_T$ factorization approach is quite similar to the $F_2^c$ case
discussed above.

The ratio $ R^c = F_L^c/F_2^c$ is 
shown in Fig. 8. We see  $ R^c \approx 0.1 \div 0.3 $ in a wide 
region of $Q^2$.
The estimation of $R^c$ is very close to the results for $R=F_L/(F_2-F_L)$
ratio (see Refs. \cite{KoPaFL}-\cite{CCFRr}).
We would like to note that these values of $ R^c $
contradict the estimation
obtained in Refs.~\cite{ZEUS2, H1c}. The effect of $ R^c $
on the corresponding differential cross-section
should be considered in the extraction of $F_2^c$
from future more precise measurements.

\begin{figure}
%\begin{center}
%\epsfig{figure=RccZEUSx.eps,width=15cm,height=15cm}
\epsfig{figure=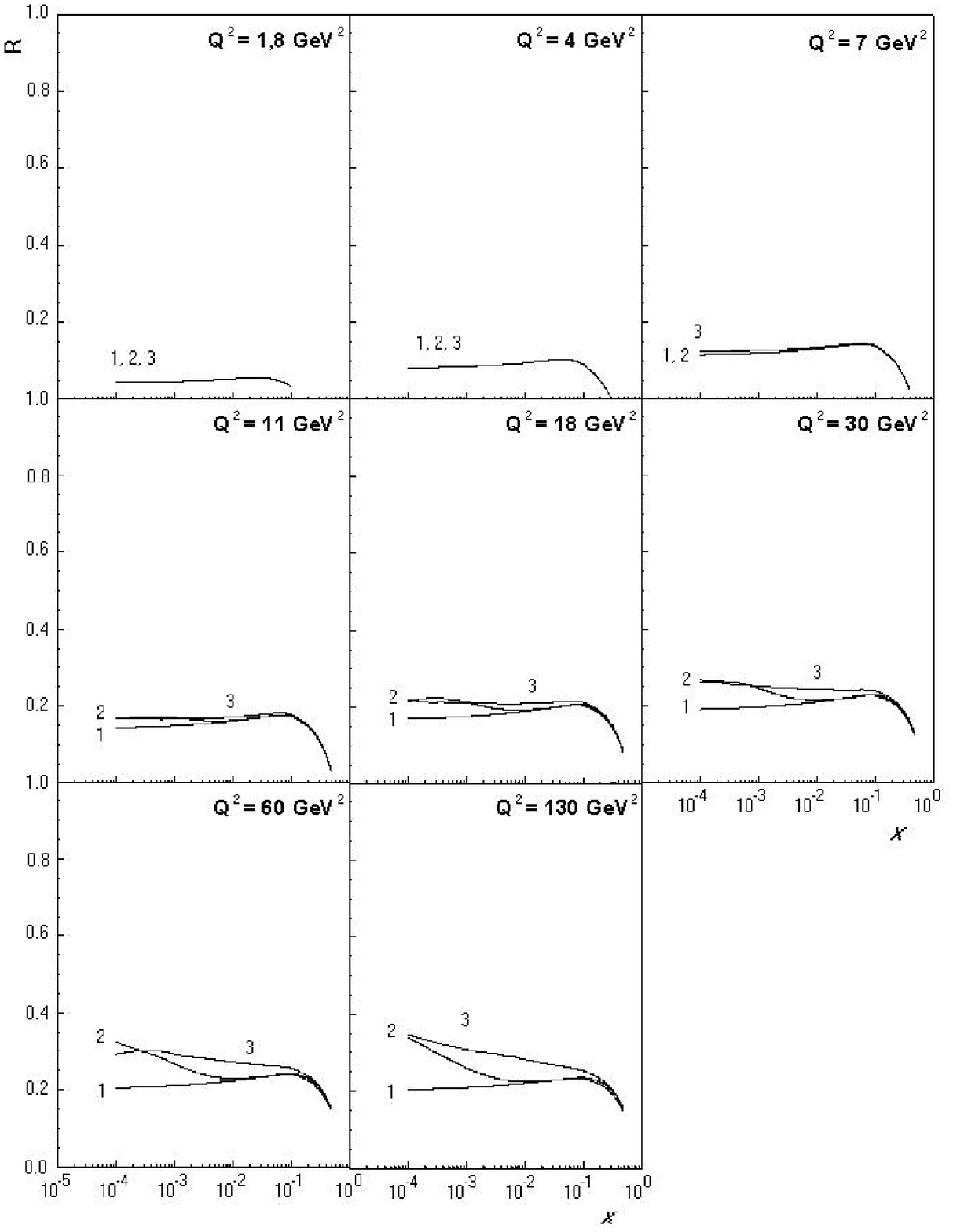,width=17cm,height=20cm}
%\end{center}
\caption{The ratio  $R^c  = F_L^c(x,Q^2)/F_2^c$ as a function of $x$ 
for different values of $Q^2$.
Curves 1, 2 and 3 are as in Fig. 2.}
\label{fig8}
\end{figure}

For the ratio $ R^c $ we found quite flat $x$-behavior at low $x$ 
in the low $Q^2$ region (see Fig. 8), where approaches based on perturbative
QCD and on $k_T$ factorization give similar predictions 
(see Fig 2, 3, 5, 6 and 7).
%$Q^2 \leq 60$ GeV$^2$.
It is in agreement with the corresponding behaviour of the ratio
$R=F_L/(F_2-F_L)$ (see Ref. \cite{KoPaFL}) at quite large values of $\Delta_P$
\footnote{At small values of $\Delta_P$, i.e. when $x^{-\Delta_P} \sim Const$,
the ratio $R$ tends to zero at $x \to 0$ (see Ref. \cite{Keller}).}
($\Delta_P > 0.2-0.3$).
The low $x$ rise of $ R^c $ at high $Q^2$ disagrees with early
calculations \cite{KoPaFL} in the framework of perturbative QCD.
It could be due to the small $x$ resummation, which is important at high $Q^2$
(see Fig 2, 3, 5, 6 and 7).
We plan to study in future this effect on $R$ in the framework
of $k_T$ factorization.\\

\section{Conclusions} \indent

We have performed the calculation of the perturbative
parts for the structure functions $F_2^c$ and $F_L^c$ for a gluon target 
having nonzero momentum squared, in the process 
of photon-gluon fusion.
The results have quite compact form for both: the Feynman gauge and a
nonsense (or BFKL-like) gluon polarizations.

We have applied the
results in the framework of $k_T$ factorization approach
to the analysis of
present data for the charm contribution to $F_2$ ($F_2^c$)
and we have given the predictions for $F_L^c$.
The analysis has been performed with
several parametrizations of unintegrated gluon
distributions (RS and BFKL) for comparison. We have found good agreement
of our results, obtained with RS and BFKL parametrizations of
unintegrated gluons distributions at $Q_0^2 = 4$ GeV$^2$,
with experimental $F_2^c$ HERA data,
except at low $Q^2$ ($Q^2 \leq 7$ GeV$^2$)
\footnote{It must be noted that the cross section of inelastic $c\bar c-$
and $b\bar b-$pair photoproduction at HERA are described by the
BFKL parametrization at a smaller value of $Q_0^2$ ($Q_0^2 = 1$ GeV$^2$)
\cite{LiZo}.}. 
We have also obtained
quite large contribution of the SF $F_L^c$ at low $x$  and
high $Q^2$ ($Q^2 \geq 30$ GeV$^2$).

We would like to note the good agreement between our results for $F_2^c$
and the ones obtained in Ref. \cite{Jung} by Monte-Carlo studies. Moreover, we
have also good agreement with fits of H1 and ZEUS data for $F_2^c$
(see recent reviews in Ref. \cite{Wolf} and references therein)
based on perturbative
QCD calculations at NLO. But unlike to these fits,
our analysis uses universal unintegrated gluon distribution, which gives
in the simplest way the main contribution to the cross-section in the
high-energy limit.

It could be also very useful to evaluate the complete $F_2$ itself and 
the derivatives of $F_2$ with respect to the 
logarithms of $1/x$ and $Q^2$ with our expressions using the unintegrated
gluons.
We are considering to present this work and also the predictions for
$F_L$ in a forthcoming article.

The consideration of the SF $F_2$ in the framework of the leading-twist
approximation of perturbative QCD (i.e. for ``pure'' perturbative QCD)
leads to very good agreement (see Ref. \cite{Q2evo} and references therein) 
with HERA
data at low $x$ and $Q^2 \geq 1.5$ GeV$^2$. The agreement improves
at lower $Q^2$ when higher twist terms are 
taken into account \cite{HT}. As
it has been studied in Refs. \cite{Q2evo,HT}, the SF $F_2$  at low 
$Q^2$ is sensitive to the
small-$x$ behavior of quark distributions.
Thus, the
analysis of $F_2$ in a broader $Q^2$ range 
should require the incorporation of parametrizations for unintegrated
quark densities, introduced recently (see Ref. \cite{Ryskin} and references
therein).

The study of the complete SF $F_L$ should also be very interesting.
The structure function $F_L$ depends strongly on the gluon distribution
(see, for example, Ref. 
\cite{CSI...}), which in turn is determined \cite{Prytz}
by the derivative $dF_2/dlnQ^2$. Thus, in the framework of perturbative
QCD at low $x$ the relation between  $F_L$, $F_2$ and
$dF_2/dlnQ^2$ could be violated by non-perturbative contributions,
which are expected to be important in the $F_L$ case (see Ref. \cite{BaCoPe}).
The application of present analysis to $F_L$ will give a ``non pure'' 
perturbative QCD predictions for the structure function that should
be compared with  
data \cite{H1FL,CCFRr} and with the ``pure'' perturbative results of 
Ref. \cite{KoPaFL}.

\vspace{1cm}
\hspace{1cm} \Large{} {\bf Acknowledgments}    \vspace{0.5cm}

\normalsize{}

We are grateful to 
Profs. V.S. Fadin, M. Ciafaloni,
I.P. Ginzburg, H. Jung, E.A. Kuraev, L.N. Lipatov and 
G. Ridolfi for useful discussions.
N.P.Z. thanks S.P. Baranov for a careful reading of the
manuscript and useful remarks.
We thank also to participants of the Workshop ``Small $x$ phenomenology''
(Lund, March 2001) for the interest in this work and discussions.

One of the authors (A.V.K.) was supported in part
by Alexander von Humboldt fellowship and INTAS  grant N366.
G.P. acknowledges the support of Xunta de Galicia
(PGIDT00 PX20615PR) and CICyT (AEN99-0589-C02-02).
N.P.Z. also acknowledge the support of Royal Swedish Academy of
Sciences.

%\setcounter{secnumdepth}{2}
%\addcontentsline{toc}{section}{APPENDIX}
%\setcounter{section}{0}
%\setcounter{subsection}{0}
%\setcounter{equation}{0}
% \def\thesection{\Alph{section}}
%\def\thesubsection {\thesection.\arabic{subsection}}
%\def\theequation{\thesection.\arabic{equation}}
%
%\appendix
%\section{Appendix}

\setcounter{secnumdepth}{2}
\addcontentsline{toc}{section}{APPENDIX}

\setcounter{section}{0}
\setcounter{subsection}{0}
\setcounter{equation}{0}
 \def\thesection{\Alph{section}}
\def\thesubsection {\thesection.\arabic{subsection}}
\def\theequation{\thesection.\arabic{equation}}

%\appendix
\section{Appendix}

Here we present the contribution to the amplitude of the DIS process
from scalar diagrams 
\footnote{These diagrams appear after calculation of the traces of
diagrams in Fig.1.}
in the elastic forward scattering of a photon on a parton.
In analogy to Eq. (\ref{2a}) one can 
represent any one-loop diagram of the elastic forward scattering 
 \begin{eqnarray}
D_{m_1m_2m_3m_4} \equiv \int \frac{d^Dk_1}{{(2\pi)}^{D/2}}
%f(k_1^2,(k-k_1)^2,(q+k_1)^2,...) 
\, d^{m_1}(k_1)\, d^{m_2}(k-k_1)\, d^{m_3}(q+k_1)\, d^{m_4}(q+k_1-k),
%~~~~~~~ \left(  d(k)=(k^2-m^2)^{-1}  \right)
 \label{a0}
 \end{eqnarray}
where $d(k)=(k^2-m^2)^{-1}$,
in the form
\footnote{This method is very similar to that in Refs.
\cite{KaKo,KoTMF} in the case of zero quark masses.
Usually the consideration of nonzero masses
into Feynman integrals complicates strongly the analysis and requires the
use of special techniques (see, for example, Ref. 
\cite{DEM}) to evaluate the diagrams. Here it
is not the case, the nonzero quark masses only modifies
the upper limit of the integral with respect to the Bjorken variable
(see the r.h.s. of Eq. (\ref{a1})).}:
\bea
D_{m_1m_2m_3m_4} ~=~
 \sum_{n=0}^{\infty} 
{\biggl(\frac{1}{x} \biggr)}^n \tilde k(n) 
\; \int_{0}^{1/(1+4a+b)}dz~ z^{n-1} 
\beta(z) \tilde f_{m_1m_2m_3m_4}(z),
%\nonumber
 \label{a1}
 \end{eqnarray}
For application of Eqs. (\ref{a0}) and (\ref{a1})
to $F_2$ and $F_L$ coefficient functions, only even $n$ are 
needed. We choose $\tilde k(n)$ so that $\tilde k(2n)=1$. \\

Below we rewrite Eq. (\ref{a1}) in the symbolic form:
 \begin{eqnarray}
D_{m_1m_2m_3m_4} ~\longrightarrow~  \tilde k(n),~  \tilde f_{m_1m_2m_3m_4}(z)
%\nonumber
 \label{a2}
 \end{eqnarray}

Then, we can represent the needed formulae by:\\

{\bf 1.} The loops:
 \begin{eqnarray}
D_{0110} &\longrightarrow&   
\tilde k(n)=1,~~  ~~~~~~ \tilde f_{0110}(z)=1,
\nonumber\\
D_{1001} &\longrightarrow&   
\tilde k(n)=(-1)^n,~~  \tilde f_{1001}(z)=1
 \label{a2l}
 \end{eqnarray}
%where 
%$d(k)={(k^2-m^2)}^{-1}$ is propagator.\\

\vskip 0.5cm

{\bf 2.} The triangles:
 \begin{eqnarray}
D_{1110} &=& D_{0111} ~\longrightarrow~   
\tilde k(n)=1,~~ ~~~~~~ \tilde f_{1110}(z)=\tilde f_{0111}(z)=-z \; f_1(z)
\nonumber\\
D_{1011} &=& D_{1101} ~\longrightarrow~   
\tilde k(n)=(-1)^n,~~  \tilde f_{1011}(z)=\tilde f_{1101}(z)=-z \; f_1(z)
 \label{a2t}
 \end{eqnarray}

\vskip 0.5cm

{\bf 3.} The boxes:

 \begin{eqnarray}
D_{2110} &\longrightarrow&   
\tilde k(n)=1,~~  ~~~~~~ ~~~~~~~~~
\tilde f_{0110}(z)= -4z^2 \; f_2(z)
%\nonumber
 \label{a2b1}\\
%& & \nonumber \\
D_{1111} &\longrightarrow&
\tilde k(n)=\frac{(1+(-1)^n)}{2},~~  
%\nonumber \\& &  
\tilde f_{1111}(z)=4z^2 \; f_1(z)
%\nonumber
 \label{a2b2}
 \end{eqnarray}

We would like to note that the result for the second box $D_{1111}$ is very 
similar to ones for the triangles. In the case of massless quarks this
property has been observed in Refs. \cite{KaKo,KoTMF}.

\setcounter{equation}{0}

\section{Appendix}

We compare the results obtained in Sect. 2
and Appendix A with well known formulae obtained in earlier works
(see Refs. \cite{BFaKh,BGMS}).

Following Ref. \cite{BGMS} let us consider the kinematics of virtual 
$\gamma \gamma$ forward scattering.
According to the optical theorem (see Sect. 1)
the quantity $F^{\mu \nu \alpha \beta}$
is the absorptive part of the  $\gamma \gamma$ forward amplitude,
connected with the cross section in the usual way. (The expression of the
amplitude in terms of the electromagnetic currents is given in Sect. 1).

In the expansion of $F^{\mu \nu \alpha \beta}$ into invariant functions
one should take into account
Lorentz invariance, $T$-invariance (symmetry in the substitution 
$\mu \nu \leftrightarrow \alpha \beta$) and gauge invariance as well, i.e.
\footnote{Sometimes we replace $q \to q_1$ and $k \to q_2$
with the purpose of keeping the symmetry $1 \leftrightarrow 2$ in our
formulae in the first part of Appendix B.} 
\bea
q_1^{\mu}F^{\mu \nu \alpha \beta} = q_1^{\nu}F^{\mu \nu \alpha \beta}=
q_2^{\alpha}F^{\mu \nu \alpha \beta}= q_2^{\beta}F^{\mu \nu \alpha \beta}
\label{B1}
\eea

The tensors in which $F^{\mu \nu \alpha \beta}$ is expanded can be constructed
in terms of vectors $q_1^{\mu}$, $q_2^{\mu}$ and the tensor $g^{\mu\nu}$.
In order to take into account explicitly gauge invariance, it is convenient 
to use their linear combinations:
\bea
{\cal Q}_1^{\mu} &=& \sqrt{\frac{-q_1^{2}}{{\cal X}}}
\biggl[q_2^{\mu} - q_1^{\mu} \frac{(q_1q_2)}{q_1^{2}} \biggr],~~
{\cal Q}_{2}^{\mu} ~=~ \sqrt{\frac{-q_2^{2}}{{\cal X}}}
\biggl[q_1^{\mu} - q_2^{\mu} \frac{(q_1q_2)}{q_2^{2}} \biggr],
\label{B2} \\
& & \nonumber \\
R^{\mu\nu} &=& R^{\nu\mu} ~=~ -g^{\mu\nu} + {{\cal X}}^{-1}
\biggl[(q_1q_2)(q_1^{\mu}q_2^{\nu} + q_1^{\nu}q_2^{\mu}) 
-q_1^{2}q_2^{\mu}q_2^{\nu} - q_2^{2}q_1^{\mu}q_1^{\nu}
\biggr],
\label{B3}
\eea
where
\bea
{\cal X} = (q_1q_2)^2 -q_1^{2}-q_2^{2}
\label{B4}
\eea

The unit vectors ${\cal Q}_i^{\mu}$ are orthogonal to the vectors 
$q_i^{\mu}$ and the symmetrical tensor $R^{\mu\nu}$ is orthogonal both to
$q_1^{\mu}$ and $q_2^{\mu}$, i.e. to ${\cal Q}_1^{\mu}$ and ${\cal Q}_2^{\mu}$:
\bea
q_1^{\mu}{\cal Q}_1^{\mu} = q_2^{\mu}{\cal Q}_2^{\mu}=0,~~
q_i^{\mu} R^{\mu\nu}=  {\cal Q}_i^{\mu}  R^{\mu\nu}=0,
\nonumber \\
{\cal Q}_1^2={\cal Q}_1^2=1,~~R^{\mu\nu}R^{\mu\nu}=2,~~
R^{\mu\nu}R^{\nu\rho}=-R^{\mu\rho}
\label{B5}
\eea

We note, that $R^{\mu\nu}$ is a metric tensor of a subspace which is
orthogonal to
$q_1^{\mu}$ and $q_2^{\mu}$. In the c.m.s. of the photons, only
two components of $R^{\mu\nu}$ are different from 0 ($R^{xx}=R^{yy}=1$).

The choice of independent tensors in which the expansion is carried out,
has a high degree of arbitrariness. We make this choice so that these tensors 
are orthogonal to each other, and the invariant functions have a simple
physical interpretation:
\bea
F^{\mu \nu \alpha \beta} &=& R^{\mu \alpha}R^{\nu \beta}F_{TT}
+ R^{\mu \alpha}{\cal Q}_2^{\nu}{\cal Q}_2^{\beta}F_{TS}
+ {\cal Q}_1^{\mu}{\cal Q}_1^{\alpha}R^{\nu \beta}F_{ST}
\nonumber \\
&+& 
{\cal Q}_1^{\mu}{\cal Q}_1^{\alpha}{\cal Q}_2^{\nu}{\cal Q}_2^{\beta}F_{SS}
+ \frac{1}{2}\biggl[R^{\mu \nu}R^{\alpha \beta} + R^{\mu \beta}R^{\nu \alpha}
- R^{\mu \alpha}R^{\nu \beta} \biggr]F_{TT}^{\tau}
\nonumber \\
&-&\biggl[R^{\mu \nu}{\cal Q}_1^{\alpha}{\cal Q}_2^{\beta}
+ R^{\mu \beta}{\cal Q}_1^{\alpha}{\cal Q}_2^{\nu}
+ (\mu \nu \leftrightarrow \alpha \beta) \biggr]F_{TS}^{\tau}
%\nonumber 
\label{B6}\\
&+& \biggl[R^{\mu \nu}R^{\alpha \beta} - R^{\mu \beta}R^{\nu \alpha}
\biggr]F_{TT}^{a}
-\biggl[R^{\mu \nu}{\cal Q}_1^{\alpha}{\cal Q}_2^{\beta}
- R^{\mu \beta}{\cal Q}_1^{\alpha}{\cal Q}_2^{\nu}
+ (\mu \nu \leftrightarrow \alpha \beta) \biggr]F_{TS}^{a}
% \label{B6}
\nonumber
\eea

The dimensionless invariant functions $F_{ab}$ defined here only depend on 
the invariants $W^2=(q_1+ q_2)^2$, $q_1^{2}$ and $q_2^{2}$. The first four
functions are expressed through the cross sections $\sigma_{ab}$ 
($a,b \equiv S, T$ for scalar 
%photons and $a,b \equiv T$ for 
and transverse photons, respectively).
The amplitudes $F_{ab}^{\tau}$ correspond to transitions with spin-flip for 
each of the photons with total helicity conservation. The last two amplitudes
are antisymmetric.\\

We would like to represent the results of Ref. \cite{BFaKh} in terms of our
functions, introduced in Section 2.

First of all, we return to the variables introduced in  Sect. 2.
Then, we have
\bea
{\cal X} ~=~ \frac{Q^4}{4x^2}\, \tilde \beta^2, ~~ 
{\cal Q}_1^{\mu} ~=~ \sqrt{\frac{4bx^2}{Q^2\tilde \beta^2}}
\biggl[q_2^{\mu} + \frac{1}{2bx} q_1^{\mu}  \biggr],~~
{\cal Q}_{2}^{\mu} ~=~ \sqrt{\frac{4bx^2}{Q^2\tilde \beta^2}}
\biggl[q_1^{\mu} + \frac{1}{2x}q_2^{\mu}  \biggr],
\label{B7} 
\eea

The results of Ref. \cite{BFaKh} have the form
\footnote{The original results of Ref. \cite{BFaKh} contain an additional
factor ${[-\pi \alpha^2/Q^2]}^{-1}$ in comparison with Eq. (\ref{B8}),
that has to do with the different normalization used in our article
(see Eq. (\ref{1})) and in Ref. \cite{BGMS}.}:
\bea
\tilde \beta F_{TT} &=& 4x \beta \biggl[
1+ 4(a-1-b)T +12bT^2 - \biggl\{ 1-8a^2x^2 + 2(2a-1-b 
\nonumber \\
& & +2bx(1-x(1+b)))T
+24b^2T^2 \biggr\} f_1 + bx^2 f_2 \biggr],
\nonumber \\
\tilde \beta F_{TS} &=& -16x \beta \biggl[
T -2x \biggl\{ ax - b(1 +2x(a-1-b))T
-6b^2T^2 \biggr\} f_1 
\nonumber \\
& & + bx^2(6a-b+6bT)T f_2 \biggr],
\nonumber \\
\tilde \beta F_{ST} &=& -16x \beta \biggl[
T -2x \biggl\{ ax - (1 +x(2a-1-b))T
-6b^2T^2 \biggr\} f_1 
\nonumber \\
& & + x^2(6a-b+6bT)T f_2 \biggr],
\nonumber \\
\tilde \beta F_{SS} &=& 64bxT^2 \beta \biggl[
2 - (1+2bx^2) f_1 - bx^2 f_2 \biggr],
\nonumber \\
\tilde \beta F_{TT}^{\tau} &=& 8x \beta \biggl[
2aT +(1-a^2)\frac{x^2}{\tilde \beta^2}+6bT^2 
+x^2 \biggl\{ 2(1+a+b)-b-2b(2a-1-b)T
\nonumber \\
& & -12b^2T^2 \biggr\} f_1 \biggr],
\nonumber \\
\tilde \beta F_{TS}^{\tau} &=& \tilde \beta F_{ST}^{\tau} ~=~
16 b^{1/2}
%\sqrt{b}
xT \beta \biggl[
2x-3T +x^2 \biggl\{ 2a-1-b+6bT \biggr\} f_1 \biggr],
\nonumber \\
\tilde \beta F_{TT}^{a} &=& 4 \beta \biggl[
x-4T + \biggl\{ 2T-x \biggr\} f_1 - bx^3 f_2 \biggr],
\nonumber \\
\tilde \beta F_{TS}^{a} &=& \tilde \beta F_{ST}^{a} ~=~
-16 b^{3/2}
%\sqrt{b}
x^2T \beta \biggl[2f_1 -  f_2 \biggr],
 \label{B8}
\nonumber
\eea
where
$$ T=\frac{x(1 -x(1+b))}{\tilde \beta^2} $$

Doing the needed projections on Eqs. (\ref{B6}) we can express the
above functions as combinations
of $f^{(1)},~ f^{(2)},~\tilde f^{(1)}$ and $\tilde f^{(2)}$
(see Sect. 2).

\bea
\tilde \beta^2 F_{SS} &=& f^{(1)},~~~~~ 
%\nonumber \\
\tilde \beta^2 F_{ST} ~=~ \tilde \beta^2 f^{(1)} + \frac{1}{2} f^{(2)}
\nonumber \\
\tilde \beta^2 F_{TS} &=&  \frac{1}{2} 
\biggl[\tilde f^{(2)} - f^{(2)} \biggr],~~~~~ 
%\nonumber \\
\tilde \beta^2 F_{TT} ~=~  \frac{1}{2} 
\biggl[\tilde \beta^2 \tilde f^{(1)} - f^{(1)} \biggr]
\label{B81}
\eea
The coefficient functions calculated in Sect. 2 can be expressed as 
combinations of $F_{AB}~~(A,B=S,T)$.

For non interacting gluons:
%\bea
%f^{(1)}~=~
%%&=& 
%-2F_{TT} + F_{TS}+ F_{ST} -\frac{1}{2}F_{SS},~~~
%%\nonumber \\
%f^{(2)} ~=~
%%&=& 
%\tilde \beta^2 \Bigl[ F_{SS}
%-2F_{TS} \Bigr]
%\label{B9} 
%\eea
%and, 
%thus,
\bea
\tilde \beta^2 C_2 &=& {\cal K} \; \Biggl[
F_{SS}+ F_{ST} -2 (F_{TT} + F_{TS}) \Biggr],
\nonumber \\
\tilde \beta^2 C_L &=& {\cal K} \; \Biggl[
F_{SS}-2 F_{TS} +4bx^2 \Bigl( F_{ST}
-2F_{TT} \Bigr) \Biggr]
\label{B9} 
\eea

For the BFKL projector:
%\bea
%\tilde f^{(1)} ~=~ F_{ST} -\frac{1}{2}F_{SS},~~~
%\tilde f^{(2)} ~=~ \tilde \beta^2 F_{SS}
%\label{B10} 
%\eea
%and, 
%thus,
\bea
\tilde \beta^4 C_{2,BFKL} &=& {\cal K} \; \Biggl[
F_{SS}+ F_{ST} +F_{TS} + F_{TT} \Biggr],
\nonumber \\
\tilde \beta^4 C_{L,BFKL} &=& {\cal K} \; \Biggl[
F_{SS}+ F_{TS} +4bx^2 \Bigl( F_{ST}
+F_{TT} \Bigr) \Biggr]
\label{B9} 
\eea

%\vskip 2cm

\setcounter{equation}{0}

%\section*{\protect\Large Appendix C}
\section{Appendix}

Here we consider the particular cases: $k^2 =0$, $m^2 =0$ and
$Q^2 \to 0$ which are relevant to compare with others.

%\vskip 2cm

\subsection{The case $k^2 =0$}

%\vskip 2cm

%{\bf a)} 
When $k^2 = 0$ 
 \begin{eqnarray}
C^g_{2}(x) ~=~
%&=& 
%e_c^2 \cdot \frac{\alpha_s(Q^2)}{4\pi}\cdot x 
{\cal K} \;
\left[
f^{(1)} + 
\frac{3}{2}\; f^{(2)} \right]
~~~\mbox{ and }~~~ 
%\nonumber \\
C^g_{L}(x) ~=~
%&=& 
%e_c^2 \cdot \frac{\alpha_s(Q^2)}{4\pi}\cdot x 
{\cal K} \;
%\left[
%4bx^2 f^{(1)} + \frac{(1+2bx^2)}{\tilde \beta^2}\cdot 
f^{(2)} 
%\right]
%\nonumber 
\label{6.1}
 \end{eqnarray}
with
\begin{eqnarray}
f^{(1)} &=& -2 \; \beta \Biggl[ \biggl(1-2x(1-x)(1-2a) \biggr) - 
\biggl(1-2x(1-2a)+2x^2(1-4a^2) \biggr) 
\nonumber \\
& & \; L(\beta)
%& &\cdot 
%ln\frac{1+\beta }{1-\beta }  
\Biggr]
%\nonumber \\
%&+&  4bx^2 ...
 \label{6.2} \\
& & \nonumber \\
f^{(2)} &=& 8x \; \beta \Biggl[(1-x) 
-2xa \;  L(\beta)
%ln\frac{1+\beta }{1-\beta }  
\Biggr],  
%\nonumber \\
%&+&  4bx^2 ...
\label{6.3} 
\end{eqnarray}
where
$$\beta^2=1-\frac{4ax}{(1-x)}$$
and
the function $L(\beta)$ has been defined in Eq. (\ref{c2.1}).

Equations (\ref{6.1})-(\ref{6.3}) 
coincide with the results of Ref. \cite{Witten}.

Indeed, we have

%{\bf a)} When $k^2 = 0$ 
 \begin{eqnarray}
C^g_{2} &=& 
%e_c^2 \cdot \beta \frac{\alpha_s(Q^2)}{4\pi}\cdot (-2x) 
{\cal K} \; (-2) \beta
\Biggl[ \biggl(1-4x(2-a)(1-x) \biggr)
\nonumber \\
&-& 
\biggl(1-2x(1-2a)+2x^2(1-6a-4a^2) \biggr) \;  L(\beta) 
%ln\frac{1+\beta }{1-\beta }  
\Biggr]
% \nonumber 
\label{6.6}\\
C^g_{L} &=& 
%e_c^2 \cdot \frac{\alpha_s(Q^2)}{4\pi}\cdot (8x^2) 
{\cal K} \; 8x
\beta \Biggl[ (1-x)
%\nonumber \\
- 2xa \;  L(\beta)
%ln\frac{1+\beta }{1-\beta }  
\Biggr]
 \label{6.7}
 \end{eqnarray}
\vskip 0.5cm

The consideration of the BFKL projector does not change the results
given above because the additional terms (see Eq. (\ref{3dd})) are
proportional to $k^2$ and they are negligible.
The expression in Eq. (\ref{6.7}) also coincides with the corresponding
result in Ref. \cite{CaCiHa2} (see Eqs. (A17,A18)).

\subsection{The case $m^2 =0$}

%\vskip 2cm

%\vskip 0.5cm
%{\bf b)} 
When $m^2 = 0$ the coefficient functions
 $C^g_k(x)$ are defined through $f^{(1)}$ and $f^{(2)}$ (see
Eqs. (\ref{3}), (\ref{4})) being in this case
 \begin{eqnarray}
f^{(1)} &=& -2 \Biggl[ 2 - 
\biggl(1-2x(1+b)+2x^2(1+b)^2 \biggr) \;  L(\tilde \beta)
%\frac{1}{\tilde \beta} \cdot
%ln\frac{1+\tilde \beta}{1-\tilde \beta}  
\Biggr]
%\nonumber \\
%&+&  ax ...
 \label{6.4} \\
& & \nonumber \\
f^{(2)} &=& 8x (1+b)(1-(1+b)x) \Biggl[1 -
2bx^2  \;  L(\tilde \beta)
%\frac{1}{\tilde \beta} \cdot
%ln\frac{1+\tilde \beta}{1-\tilde \beta}  
\Biggr]
%\nonumber \\
%&+&  ax ...
\label{6.5}
\end{eqnarray}

For the coefficient functions themselves, we have

%{\bf b)} 
%When $m^2 = 0$
 \begin{eqnarray}
\tilde \beta^4 \; C^g_{2} &=& 
%e_c^2 \; \frac{\alpha_s(Q^2)}{4\pi}\cdot (-2x) 
{\cal K} \; (-2)
\left(1-x(1+b)\right)
\Biggl[ 2 
\biggl(1-2x(1+b)+\frac{x^2(1-b)^2}{1-x(1+b)} \biggr)
\nonumber \\
&-& \biggl(1-x(1+b)-4x^3b(1+b) +\frac{x^2(1-b)^2}{1-x(1+b)} \biggr)
 \;  L(\tilde \beta)
%\frac{1}{\tilde \beta} \cdot
%ln\frac{1+\tilde \beta}{1-\tilde \beta}  
\Biggr]
 \label{6.8}\\
\tilde \beta^4 \; C^g_{L} &=& 
%e_c^2 \cdot \frac{\alpha_s(Q^2)}{4\pi}\cdot (8x) 
{\cal K} \; 8x
\left(1-x(1+b)\right)
\Biggl[ \biggl((1+b)-2bx\biggl[1+ \frac{x^2(1-b)^2}{1-x(1+b)} \biggr]
\biggr)
\nonumber \\
&+& bx\biggl(1-3x(1+b)+4x^3b(1+b) + \frac{x^2(1-b)^2}{1-x(1+b)} \biggr)
 \;  L(\tilde \beta)
%\frac{1}{\tilde \beta} \cdot
%ln\frac{1+\tilde \beta}{1-\tilde \beta}  
\Biggr]
 \label{6.9}
 \end{eqnarray}

%{\bf b)} When $m^2 = 0$  
In the case of the BFKL projector,
the coefficient functions
 $C^g_k(x)$ are defined by Eqs. (\ref{6.1}), (\ref{5}) and  (\ref{6})
with the replacement $f^{(i)} \to f^{(i)}_{BFKL}$ as in 
Eq. (\ref{4dd}). In Eq. (\ref{4dd}) the expressions for $f^{(i)}$ 
can be found in Eqs. (\ref{6.4}), (\ref{6.5}) while for
$ \tilde f^{(i)}$ are given by:

 \begin{eqnarray}
\tilde f^{(1)} &=& - \frac{(1+b)(1-x(1+b))}{bx}
\Biggl[1 -
2bx^2  \;  L(\tilde \beta)
%\frac{1}{\tilde \beta} \cdot
%ln\frac{1+\tilde \beta}{1-\tilde \beta} 
 \Biggr]
= - \frac{1}{8bx^2} f^{(2)}
%\nonumber \\
%&+&  ax ...
 \label{6.4dd} \\
& & \nonumber \\
\tilde f^{(2)} &=& 4 (1-x(1+b))^2
\Biggl[ 3 - (1+2bx^2)  \;  L(\tilde \beta)
%\frac{1}{\tilde \beta} \cdot
%ln\frac{1+\tilde \beta}{1-\tilde \beta}  
\Biggr]
%\nonumber \\
%&+&  ax ...
\label{6.5dd}
\end{eqnarray}
and, thus,
 \begin{eqnarray}
f^{(1)}_{BFKL} &=& 
C^g_{2}/{\cal K} 
%\biggl(
%e_c^2 \cdot \frac{\alpha_s(Q^2)}{4\pi}\cdot x \biggr)
 \label{6.4d3} \\
& & \nonumber \\
\tilde \beta^4 \; f^{(2)}_{BFKL} &=& 8x (1-x(1+b))
\Biggl[ 1+b-18bx(1-x(1+b)) \nonumber \\
&+& 2bx 
\left(3-4x(1+b)+6bx^2(1-x(1+b)) \right)  \;  L(\tilde \beta)
%\frac{1}{\tilde \beta} \cdot
%ln\frac{1+\tilde \beta}{1-\tilde \beta}  
\Biggr]
%\nonumber \\
%&+&  ax ...
\label{6.5d3}
\end{eqnarray}

For the coefficient functions $C^g_{k,BFKL}(x)$ we have the following
results:
 \begin{eqnarray}
\tilde \beta^8 \; C^g_{2,BFKL} &=& 
%e_c^2 \cdot \frac{\alpha_s(Q^2)}{4\pi}\cdot (-2x) 
{\cal K} \; (-2)
\left(1-x(1+b)\right)
\Biggl[ 2 
\biggl(1-5x(1+b) \nonumber \\
&+& x^2(1+48b+b^2) + x^3(1+b)(1-48b+b^2) +
\frac{x^4(1-b)^4}{1-x(1+b)} \biggr)
\nonumber \\
&-& \biggl(1-x(1+b)+(1-30b+b^2)x^2 +(1-50b+b^2)x^3(1+b)
\nonumber \\
&+& 72x^4b^2
-56x^5b^2(1+b) +\frac{x^4(1-b)^4}{1-x(1+b)} \biggr)
 \;  L(\tilde \beta)
%\frac{1}{\tilde \beta} \cdot
%ln\frac{1+\tilde \beta}{1-\tilde \beta}  
\Biggr]
 \label{6.8}\\
\tilde \beta^8 \; C^g_{L,BFKL} &=& 
%e_c^2 \cdot \frac{\alpha_s(Q^2)}{4\pi}\cdot (8x) 
{\cal K} \; 8x
\left(1-x(1+b)\right)
\Biggl[ \biggl((1+b)-20bx + 24b(1+b)x^2
\nonumber \\
&-&
2b(1+12b+b^2)x^3-
2(1+b)b(1-12b+b^2)x^4 -
2bx \; \frac{x^4(1-b)^4}{1-x(1+b)} 
\biggr)
\nonumber \\
&+& bx\biggl(7-11x(1+b)+(1-42b+b^2)x^2 +(1-30b+b^2)(1+b)x^3 
\nonumber \\
&+&24b^2x^4
-8b^2(1+b)x^5
 + \frac{x^4(1-b)^4}{1-x(1+b)} \biggr)
 \;  L(\tilde \beta)
%\frac{1}{\tilde \beta} \cdot
%ln\frac{1+\tilde \beta}{1-\tilde \beta}  
\Biggr]
 \label{6.9d}
 \end{eqnarray}

\subsection{The case $Q^2\to 0$}

%\vskip 2cm
%{\bf b)} 
Using the definitions in Eq. (\ref{c1}),
when $x \to 0$ we have got the following relations (at $O(x)$):

for the intermediate functions:
\bea 
\tilde \beta^2 &=& 1-4x\Delta, ~~  
\beta^2 = \hat \beta^2 
%\left(1+\frac{x \rho}{(1-\Delta)(1-\Delta-\rho)} \right)
%\equiv \hat \beta^2
(1-4\gamma x), \nonumber \\
f_1&=& L(\hat \beta)\left(1+2x(\gamma+\Delta)\right)
-2x(\gamma+\Delta)\frac{(1-\Delta)}{z}
,~~ \nonumber \\
f_2&=&-\frac{2(1-\Delta)}{z} \biggl[1-
2x\frac{1-\Delta-2z}{z}\; (\gamma+\Delta) \biggr], 
 \label{c4.1}
\eea
where
$$ z=\frac{\rho}{2},~~~ \gamma=\frac{z/2}{(1-\Delta)(1-\Delta-2z)}$$

for the basic functions:
\bea
f^{(1)} &=& -2 \hat \beta 
\Biggl[ 1-2(1-\Delta)(\Delta-z)-
\Bigl(1-2(1-\Delta)\Delta +2(1-z)z \Bigr) 
L(\hat \beta) 
\nonumber \\
&+& 2x \Biggl\{
\frac{1-\Delta }{z} \left(\Delta +(\Delta-z)(1-2z\Delta) \right) -
\frac{1-\Delta-z }{z}\,\gamma \nonumber \\
&+&
\biggl(1-\Delta \Bigl[3-2(1-\Delta)\Delta +2(1-z)z\Bigr] \biggr) 
L(\hat \beta) \Biggr\} \Biggr],
 \label{c4.2}\\
f^{(2)} &=& 8x \hat \beta 
\Biggl[ (1-\Delta) \left(1-2\frac{\Delta (1-\Delta)}{z}(z-\Delta) \right)
- \Bigl(2\Delta ^2 (1-\Delta) 
\nonumber \\
&+& z(1-2\Delta (1-\Delta)) \Bigr)
L(\hat \beta) \Biggr],
\nonumber \\
x\tilde f^{(1)} &=& - \hat \beta \Biggl[ \biggl\{(1-\Delta)-x
(3-6\Delta+4\Delta^2) \biggr\} - \biggl\{z+2x
(\Delta(1-\Delta)-z) \biggr\}
 L(\hat \beta)
\Biggr],
\nonumber \\
x\tilde f^{(2)} &=& 4x \hat \beta (1-\Delta)^2 \Bigl[2- L(\hat \beta)
\Bigr],
 \label{c4.3}
\eea
and, thus, 
\bea
f^{(1)}_{BFKL} &=& f^{(1)} ~+~ 4 \hat \beta \Delta
\Biggl[ 3\Bigl(1-\Delta -z L(\hat \beta) \Bigr) 
%\nonumber \\
%&-& 
-x \Biggl\{
11-46\Delta +40\Delta^2 +4z(1-\Delta)  
\nonumber \\
&-& 2 \Bigl[1-5(1-\Delta)\Delta +z(5-2z-12\Delta)\Bigr]  
\Biggr\} L(\hat \beta)  \Biggr],
%
%-2 \hat \beta 
%\Biggl[ 1-2(1-\Delta)(4\Delta-z)-
%\left(1-2(1-\Delta)\Delta +2z(1-z -3\Delta) \right) 
%L(\hat \beta) 
%\nonumber \\
%&+& 2x \Biggl\{
%\frac{1-\Delta-z }{z}\,(2\Delta-\gamma) -1  
%+ 2\Delta \left[7-24\Delta +21\Delta^2 +3z(1-\Delta) \right] 
%\nonumber \\
%&+&
%\left(1-\Delta \left[5-12(1-\Delta)\Delta +6z(2-z-4\Delta)\right] \right) 
%L(\hat \beta) \Biggr\} \Biggr],
 \label{c4.4}\\
f^{(2)}_{BFKL} &=& f^{(2)} ~-~
48x \hat \beta \Delta (1-\Delta) \Bigl[2- L(\hat \beta)
\Bigr].
%8x \hat \beta 
%\Biggl[ (1-\Delta) \left(1-2\frac{\Delta (1-\Delta)}{z}(7z-\Delta) \right)
%+ (2\Delta (1-\Delta)(3-4\Delta) 
%\nonumber \\
%&-& z(1-2\Delta (1-\Delta)) )
%L(\hat \beta) \Biggr],
 \label{c4.5}
\eea

Similarly to Eq. (\ref{c4.3}),  the coefficient functions in Eq. (\ref{c3.2})
and the functions $f^{(1)}_{BFKL}$ in Eq. (\ref{c3.3})
have the additional terms proportional $x$.

\bea
C^g_{2}/{\cal K}  &=& f^{(1)} ~+~ 4 x\hat \beta 
\Biggl[\,6 \,\frac{\Delta ^2(1-\Delta)^2 }{z}
+3-\Delta \Bigl(
11-16\Delta +10\Delta^2 +4z(1-\Delta) \Bigr)  
\nonumber \\
&+& \Biggl\{
2 \Delta \Bigl[1-5(1-\Delta)\Delta +z(5-2z-3\Delta)\Bigr] -3z \Biggr\} 
L(\hat \beta)  \Biggr],
%%e_c^2 \cdot \frac{\alpha_s(Q^2)}{4\pi}
%{\cal K} \cdot (-2 \hat \beta) 
%\Biggl[ 1-2(1-\Delta)(\Delta-z)-
%\biggl(1-2(1-\Delta)\Delta 
%\nonumber \\
%&+& 2z(1-z)\biggr) 
%L(\hat \beta) 
%+ 2x \Biggl\{
%\frac{1-\Delta-z }{z}\,(2\Delta (1-3\Delta(1-\Delta))-\gamma) +2  
%\nonumber \\
%&+& 2\Delta \left[7-12\Delta +9\Delta^2 +3z(1-\Delta) \right] 
%\nonumber \\
%&+&
%\left(1+3z -\Delta \left[5-12\Delta(1-\Delta) +6z(2-z-\Delta)
%\right] \right) 
%L(\hat \beta) \Biggr\} \Biggr],
 \label{c4.6}\\
C^g_{L}/{\cal K} &=& 
%e_c^2 \cdot \frac{\alpha_s(Q^2)}{4\pi}
%{\cal K} \cdot 
8x \hat \beta 
\Biggl[\, 2\, \frac{\Delta^2(1-\Delta)^2}{z} +
1-2\Delta \Bigl(
2-3\Delta +2\Delta^2 +z(1-\Delta) \Bigr)  
\nonumber \\
&+& \Biggl\{
 \Delta \Bigl[1-4(1-\Delta)\Delta +2z(2-z-\Delta)\Bigr] -z \Biggr\} 
L(\hat \beta) \Biggr],
 \label{c4.7}
\eea

\bea
C^g_{2,BFKL}/{\cal K} &=& C^g_{2}/{\cal K} ~+~ 4 \hat \beta \Delta
\Biggl[ 3\Bigl(1-\Delta -z L(\hat \beta) \Bigr) 
%\nonumber \\
%&-& 
-x \Biggl\{
47-130\Delta +88\Delta^2   
\nonumber \\
&+&4z(1-\Delta)
- 2 \Bigl[10-23\Delta +14\Delta^2 +z(5-2z-18\Delta)\Bigr]  L(\hat \beta)
\Biggr\}   \Biggr],
%%e_c^2 \cdot \frac{\alpha_s(Q^2)}{4\pi}
%{\cal K} \cdot (-2 \hat \beta) 
%\Biggl[ 1-2(1-\Delta)(4\Delta-z)-
%\biggl(1-2(1-\Delta)\Delta 
%\nonumber \\
%&+& 2z(1-z -3\Delta) \biggr) 
%L(\hat \beta) 
%%\nonumber \\&+& 
%+2x \Biggl\{
%\frac{1-\Delta-z }{z}\,(2\Delta (1-3\Delta(1-\Delta))-\gamma) -4  
%\nonumber \\
%&+& \Delta \left[61-154\Delta +106\Delta^2 +10z(1-\Delta) \right] 
%\nonumber \\
%&+&
%\left(1+3z -\Delta \left[25-58\Delta+40\Delta^2 +2z(11-5z-21\Delta)
%\right] \right) 
%L(\hat \beta) \Biggr\} \Biggr],
 \label{c4.8}\\
C^g_{L,BFKL}/{\cal K} &=& C^g_{L}/{\cal K} ~-~ 48x \hat \beta \Delta
\Biggl[ (1-\Delta)(2-3\Delta) 
%\nonumber \\
%&-&  
-\Bigl[(1-\Delta)^2 -z\Delta\Bigr]  
 L(\hat \beta)  \Biggr]
%%e_c^2 \cdot \frac{\alpha_s(Q^2)}{4\pi}
%{\cal K} \cdot 8x \hat \beta 
%\Biggl[ 2 \frac{\Delta^2(1-\Delta)^2}{z} +
%1-2\Delta -2 \Delta(1-\Delta)
% \left[7-11\Delta +z\right]
%\nonumber \\
%&-& \left(\Delta (5-8\Delta +2\Delta^2 -2z(1-z))
%+z(1-2\Delta +8\Delta^2) \right)
%L(\hat \beta) \Biggr],
 \label{c4.9}
\eea

%\newpage

%\vspace{0.5cm}

%\hspace{1cm} {\Large{\bf Figure captions}}    \vspace{0.5cm}

\end{document}